\documentclass[11pt]{article}

\usepackage{times}
\usepackage{fullpage}
\usepackage{amsmath}
\usepackage{amsfonts}
\usepackage{supertabular}
\usepackage{graphicx}
\usepackage{subfigure}
\usepackage{epsfig}
\usepackage{xspace}

\def\asinh{{\rm asinh}}

\newcommand{\mbs}[1]{\boldsymbol{#1}}

  \def\bC{{\mbs{C}}}
  \def\bF{{\mbs{F}}}

 \def\bQ{{\mbs{Q}}}

 \def\bb{{\mbs{b}}} \def\bc{{\mbs{c}}}

\def\bm{{\mbs{m}}}  
  
\def\bs{{\mbs{s}}}

\def\bcc{bcc\xspace}
\def\fcc{fcc\xspace}

\title{A Micromechanical Model of Hardening, Rate sensitivity and
Thermal Softening in BCC Single Crystals}

\author{L. Stainier \\
        Laboratoire de Techniques A\'eronautiques et Spatiales \\
        University of Li\`ege \\        4000 Li\`ege, Belgium \\[.5em]
        \and A.~M. Cuiti\~no \\
        Department of Mechanical and Aerospace Engineering \\
        Rutgers University \\
        Piscataway, NJ 08854, USA \\[.5em]
        \and M. Ortiz \\
        Graduate Aeronautical Laboratories \\
        California Institute of Technology \\
        Pasadena, CA 91125, USA
        }

\begin{document}
\maketitle

\begin{abstract}

The present paper is concerned with the development of a
micromechanical model of the hardening, rate-sensitivity and thermal
softening of \bcc crystals.  In formulating the model we specifically
consider the following unit processes: double-kink formation and
thermally activated motion of kinks; the close-range interactions
between primary and forest dislocations, leading to the formation of
jogs; the percolation motion of dislocations through a random array of
forest dislocations introducing short-range obstacles of different
strengths; dislocation multiplication due to breeding by double
cross-slip; and dislocation pair annihilation.  The model is found to
capture salient features of the behavior of Ta crystals such as: the
dependence of the initial yield point on temperature and strain rate;
the presence of a marked stage I of easy glide, specially at low
temperatures and high strain rates; the sharp onset of stage II
hardening and its tendency to shift towards lower strains, and
eventually disappear, as the temperature increases or the strain rate
decreases; the parabolic stage II hardening at low strain rates or
high temperatures; the stage II softening at high strain rates or low
temperatures; the trend towards saturation at high strains; the
temperature and strain-rate dependence of the saturation stress; and
the orientation dependence of the hardening rate.

\end{abstract}

\newpage

\section{Introduction}

The present paper is concerned with the development of a
micromechanical model of the hardening, rate-sensitivity and thermal
softening of \bcc crystals.  We place primary emphasis on the
derivation of closed-form analytical expressions describing the
macroscopic behavior of the crystals amenable to implementation as
constitutive relations within a standard finite-element code.  In
developing the model, we follow the well-established paradigm of
micromechanical modeling, consisting of: the identification of the
dominant or rate-limiting `unit' processes operating at the
microscale; the identification of the macroscopic forces driving the
unit processes; the analysis of the response of the unit processes to
the macroscopic driving forces; and the determination of the average
or macroscopic effect of the combined operation of all the
micromechanical unit processes (see, e.~g., \cite{BulatovKubin1998,
Phillips1998, CampbellFoilesHuang1998, PhillipsRodneyShenoy1999,
MoriartyXuSoderlind1999, Baskes1999} for recent overviews and
discussions pertaining to micromechanics and multiscale modeling of
materials).

In formulating the present model we specifically consider the
following unit processes: double-kink formation and thermally
activated motion of kinks; the close-range interactions between
primary and forest dislocations, and the subsequent formation of
jogs; the percolation motion of dislocations through a random
array of forest dislocations introducing short-range obstacles of
different strengths; dislocation multiplication due to breeding by
double cross-slip; and dislocation pair annihilation. We believe
that this forms an `irreducible' set of unit processes, in that
each of these processes accounts for--and is needed for
matching--salient and clearly recognizable features of the
experimental record. For instance, consideration of dislocation
multiplication and annihilation leads to a predicted stage I-II
transition strain which decreases with temperature and increases
with strain rate, as observed experimentally.

We bring an assortment of analysis tools to bear on each of the
unit processes under consideration. As already mentioned, the
choice of tools is to a large extent conditioned by our desire to
derive closed-form analytical expressions for all constitutive
relations. The motion of dislocations through an otherwise
defect-free lattice is assumed to be thermally activated and
controlled by the nucleation of kink pairs. A analysis of this
process based on transition-state theory yields the effective
Peierls stress of the lattice as a function of temperature and
strain rate. The short-range interactions between a primary and
secondary dislocation are assumed to result in the acquisition by
both dislocations of a jog. The process by which the primary
dislocation unpins from a forest obstacle is assumed to be
thermally activated and the activation energy barrier is
identified with the jog-formation energy. These assumptions result
in temperature and rate-dependent obstacle strengths which are in
good qualitative and quantitative agreement with experiment.

The percolation motion of dislocations through random arrays of
point obstacles is studied in detail using statistical methods.
The analysis presented here generalizes the analysis of Cuiti\~no
and Ortiz \cite{cuitino:1992} so as to account for multi-species
distributions of finite-strength obstacles. Finally, we model
dislocation multiplication as the result of two competing effects:
the proliferation of dynamic sources by double cross-slip; and
dislocation pair annihilation. We model cross-slip as a thermally
activated process and develop an elastic model, similar to that of
Huang {\it et al.} \cite{HuangGhoniemDelaRubia1999}, which
estimates the probability that the trajectories of two parallel
screw dislocations collide, resulting in the annihilation of the
pair. The model predicts that multiplication and annihilation
balance out at sufficiently large strains, leading to saturation.

We validate the micromechanical model by recourse to detailed
comparisons with the uniaxial tension tests on Ta single crystals
of Mitchell and Spitzig \cite{mitchell:1965}. The model is found
to capture salient features of the behavior of Ta crystals such
as: the dependence of the initial yield point on temperature and
strain rate; the presence of a marked stage I of easy glide,
specially at low temperature and high strain rates; the sharp onset
of stage II hardening and its tendency to shift towards lower
strains, and eventually disappear, as the temperature increases or
the strain rate decreases; the parabolic stage II hardening at low
strain rates or high temperatures; the stage II softening at high
strain rates or low temperatures; the trend towards saturation at
high strains; the temperature and strain-rate dependence of
the saturation stress; and the orientation dependence of the
hardening rate.

\section{General Framework}
\label{GeneralFramework}

Our objective is to formulate a model of the hardening of \bcc
crystals which is well-suited to finite-deformation large-scale
finite-element calculations and, therefore, we couch the model within
a nonlinear kinematics framework. To this end, we adopt a conventional
multiplicative elastic-plastic kinematics of the form \cite{Lee1969,
Teodosiu1970, AsaroRice1977, Havner1973, HillRice1972, Rice1971}
\begin{equation}
\bF = \bF^e \bF^p \label{FeFp}
\end{equation}
where $\bF$ is the deformation gradient, $\bF^e$ is the elastic
lattice distortion and rotation, and $\bF^p$ is the plastic
deformation, which represents the net effect of crystallographic
slip and leaves the lattice undistorted and unrotated. In
addition, the plastic deformation is volume preserving. In
materials such as metals, the elastic response is ostensibly
independent of the internal processes and the free energy
density decomposes additively as
\begin{equation}
A = W^e(\bF^e, T) + W^p(T, \bF^p,\bQ) \label{PhiAdditive}
\end{equation}
where $T$ is the absolute temperature and $\bQ$ is some suitable
set of internal variables. The function $W^e$ determines the
elastic response of the metal, e.~g., upon unloading, whereas the
function $W^p$ describes the hardening of the crystal. Physically,
$W^p$ measures the energy stored in the crystal due to the plastic
working of the material.

The examples of validation reported subsequently probe the
hardening behavior of the material due to crystallographic slip
over a range of moderate temperatures and strain rates. We
therefore eschew a number of issues which become important at high
pressures and elevated temperatures, but which play a limited role
otherwise, including the volumetric equation of state (EoS), the
pressure dependence of yield and elastic moduli, and others.
First-principles calculations of the EoS and elastic moduli of
\bcc metals up to high pressures and temperatures may be found in
\cite{WassermanStixrudeCohen1996, CohenStixrudeWasserman1997,
SoderlindMoriarty1998, Steinle-neumannStixrudeCohen1999,
BulatovRichmondGlazov1999, CohenGulserenHemley2000}.  These
results may be used as a basis for extending the present theory to
conditions such as arise in shocked materials.  These extensions
notwithstanding, a simple form of the elastic energy density
appropriate for present purposes is
\begin{equation}
W^e(\bF^e,T) = \frac{1}{2} c_{ijkl}(T) (\epsilon_{ij}^e -
\alpha_{ij} T) (\epsilon_{kl}^e - \alpha_{kl} T)
\label{SimpleElasticPotential}
\end{equation}
where $\bc$ is the tensor of elastic moduli,
\begin{equation}\label{EE}
\mbs{\epsilon}^e = \log \sqrt{\bC^e}
                 = \frac{1}{2}\log({\bF^{e}}^{T}\bF^{e})
\end{equation}
is the logarithmic elastic strain, and $\mbs{\alpha}$ is the thermal
expansion tensor. The elastic properties of a cubic crystal are fully
described by the three Voigt constants $c_{11}$, $c_{12}$ and $c_{44}$
and a scalar thermal expansion coefficient $\alpha$. For low
temperatures, the temperature dependence of the moduli may be taken to
be linear in $T$ to a first approximation, leading to the relation:
\begin{equation}
c_{ijkl}(T) \approx c_{ijkl}^0 - T \; c_{ijkl}^1
\end{equation}
This linear dependence of the elastic moduli is observed
experimentally up to moderate temperatures \cite{simmons:1971},
and can be justified within the framework of statistical mechanics
\cite{weiner:1983}.

Plastic deformations in single crystals are crystallographic in
nature. Following Rice \cite{Rice1971}, we adopt a flow rule of the
form
\begin{equation}
\dot{\bF}^p = \left(\sum_{\alpha=1}^N {\dot\gamma}^\alpha
\bs^\alpha \otimes \bm^\alpha\right) \bF^p
\label{FlowRuleCrystals}
\end{equation}
where $\gamma^\alpha$ is the slip strain, and $\bs^\alpha$, and
$\bm^\alpha$ are orthogonal unit vectors defining the slip
direction and slip-plane normal corresponding to slip system
$\alpha$.  The collection $\mbs{\gamma}$ of slip strains may be
regarded as a subset of the internal variable set $\bQ$.  A zero
value of a slip rate ${\dot\gamma}^\alpha$ signifies that the
corresponding slip system $\alpha$ is inactive. The flow rule
(\ref{FlowRuleCrystals}) allows for multiple slip, i.~e., for
simultaneous activity on more than one system over a region of the
crystal.  The vectors $\{\bs^\alpha, \bm^\alpha\}$ remain constant
throughout the deformation and are determined by crystallography.
For \bcc crystals, we consider the 24 slip systems listed in
Table~\ref{tab:slip-systems} in the Schmid and Boas nomenclature.

Slip on the $\{112\}$ systems is known to be asymmetric at low
temperatures: slip is easier when the applied stress is such that
dislocations move in the twinning direction (e.~g.,
\cite{hull:1984, BengusDolginTabachnikova1985,
SeegerHollang2000}). However, over the temperature range of
interest here the experimental evidence \cite{SeegerHollang2000,
franciosi:1983} suggests that the extent of the
twinning/anti-twinning asymmetry is relatively small, e.~g., of
the order of 20 MPa for Mo at 150 K \cite{SeegerHollang2000}, and,
for simplicity, we will neglect it to a first approximation.

\begin{table}
\centerline{$
  \begin{array}{|l|c|c||l|c|c|c|}
    \hline
    \text{Syst.} & \text{Direction} & \text{Plane} &
        \text{Syst.} & \text{Direction} & \text{Plane} & \text{Group}
\\
    \hline\hline
    A2 & [\bar{1}11] & (0\bar{1}1) &
    A2'  & [\bar{1}11] & (      211) & \text{A} \\
    A3 & [\bar{1}11] & (      101) &
    A3'  & [\bar{1}11] & (12\bar{1}) & \text{T} \\
    A6 & [\bar{1}11] & (      110) &
    A6'  & [\bar{1}11] & (1\bar{1}2) & \text{T} \\
    B2 & [      111] & (0\bar{1}1) &
    B2'' & [      111] & (\bar{2}11) & \text{A} \\
    B4 & [      111] & (\bar{1}01) &
    B4'  & [      111] & (1\bar{2}1) & \text{A} \\
    B5 & [      111] & (\bar{1}10) &
    B5'  & [      111] & (11\bar{2}) & \text{A} \\
    C1 & [11\bar{1}] & (0\bar{1}1) &
    C1'  & [11\bar{1}] & (2\bar{1}1) & \text{T} \\
    C3 & [11\bar{1}] & (      101) &
    C3'' & [11\bar{1}] & (\bar{1}21) & \text{T} \\
    C5 & [11\bar{1}] & (\bar{1}10) &
    C5'' & [11\bar{1}] & (      112) & \text{A} \\
    D1 & [1\bar{1}1] & (0\bar{1}1) &
    D1'' & [1\bar{1}1] & (21\bar{1}) & \text{T} \\
    D4 & [1\bar{1}1] & (\bar{1}01) &
    D4'' & [1\bar{1}1] & (      121) & \text{A} \\
    D6 & [1\bar{1}1] & (      110) &
    D6'' & [1\bar{1}1] & (\bar{1}12) & \text{T} \\
   \hline
  \end{array}
$} \caption{Slip systems of \bcc crystals}\label{tab:slip-systems}
\end{table}

In the examples presented in Section~\ref{ComparisonExperiment},
the constitutive equations are integrated in time using the
variational update of Ortiz and Stainier \cite{OrtizStainier1999}.
The variational formulation of the rate problem proposed by Ortiz
and Stainier \cite{OrtizStainier1999} additionally furnishes a
convenient avenue for the superposition of the various unit
mechanisms analyzed subsequently. Our working assumption is that
the dissipation rates arising from these mechanisms are
\emph{additive}. This assumption in turn implies that the critical
resolved shear stress $\tau_c^\alpha$ for the operation of system
$\alpha$ may be computed as the sum of unit-process contributions,
i.~e.,
\begin{equation}\label{CRSS}
\tau_c^\alpha = \sum_{\rm processes} \tau_c^\alpha(\rm process)
\end{equation}
where the sum extends to all unit processes. In the present model,
these contributions are the Peierls stress of the lattice and the
forest-obstacle resistance. Other dissipation mechanisms, such as
phonon drag, may in principle be superposed likewise.

\section{Dislocation mobility}
\label{DislocationMobility}

In this section we begin by considering the thermally activated
motion of dislocations within an \emph{obstacle-free} slip plane.
Under these conditions, the motion of the dislocations is driven
by an applied resolved shear stress $\tau$ and is hindered by the
lattice resistance, which is weak enough that it may be overcome
by thermal activation. The lattice resistance is presumed to be
well-described by a Peierls energy function, which assigns an
energy per unit length to dislocation segments as a function of
their position on the slip plane.

In \bcc crystals, the core of screw dislocation segments relaxes
into low-energy non-planar configurations
\cite{DuesberyVitekBowen1973, Vitek1976, Vitek1992,
XuMoriarty1996, DuesberyVitek1998, MoriartyXuSoderlind1999,
Ismail-beigiArias2000, WangStrachanCaginGoddard2000}.  This
introduces deep valleys into the Peierls energy function aligned
with the Burgers vector directions and possessing the periodicity
of the lattice.  At low temperatures, the dislocations tend to
adopt low-energy configurations and, consequently, the dislocation
population predominantly consists of long screw segments. In order
to move a screw segment normal to itself, the dislocation core
must first be constricted, which requires a substantial supply of
energy. Thus, the energy barrier for the motion of screw segments,
and the attendant Peierls stress, may be expected to be large, and
the energy barrier for the motion of edge segments to be
comparatively smaller. For instance, Duesbery and Xu
\cite{DuesberyXu1998} have calculated the Peierls stress for a
rigid screw dislocation in Mo to be 0.022$\mu$, where $\mu$ is the
$\langle 111 \rangle$ shear modulus, whereas the corresponding
Peierls stress for a rigid edge dislocation is 0.006$\mu$, or
about one fourth of the screw value. This suggests that the
rate-limiting mechanism for dislocation motion is the thermally
activated motion of kinks along screw segments (\cite{Hirsch1960,
SeegerSchiller1962, hirth:1968}).

Consider now a screw dislocation segment, possibly pinned at both
ends by an obstacle pair, subjected to a resolved shear stress
$\tau > 0$.  At sufficiently high temperatures a double-kink may
be nucleated with the assistance of thermal activation (e.~g.,
\cite{HirthHoagland1993, XuMoriarty1998, MoriartyXuSoderlind1999},
and the subsequent motion of the kinks causes the screw segment to
effectively move forward, Fig.~1.  Neglecting
entropic effects and invoking Peach-Koehler's formula, the forward
and backward activation enthalpies $\Delta G^+$ and $\Delta G^-$,
respectively, for the formation of a double kink may be
approximated as
\begin{equation}
\Delta G^\pm \approx E^{\rm kink} \mp b \tau L^{\rm kink} l_P
\end{equation}
where $E^{\rm kink}$ is the energy of formation of a kink-pair,
$L^{\rm kink}$ is the length of an incipient double kink, and
$l_P$ is the distance between two consecutive Peierls valleys. For
\bcc crystals, $l_P = \sqrt{2/3} a$ if the slip plane is
$\{110\}$, $l_P = \sqrt{2} a$, if the slip plane is $\{112\}$, and
$l_P = \sqrt{8/3} a$ if the slip plane is $\{123\}$, where $a$ is
the cubic lattice size \cite{SeegerHollang2000}.

The formation energy $E^{\rm kink}$ cannot be reliably estimated
from elasticity since is is composed mostly of core region. It
can, however, be accurately computed by recourse to atomistic
models. For instance, for Mo at zero stress Xu and Moriarty
\cite{XuMoriarty1998} have found formation energies $E^{\rm kink}$
of the order of 1 eV for kinks separated by a distance greater
than $L^{\rm kink} = 15 b$. The core structure, gamma surfaces,
Peierls stress, and kink-pair formation energies associated with
the motion of $a/2 \langle 111 \rangle$ screw dislocations in Ta
and Mo \cite{MoriartyXuSoderlind1999} have also been calculated by
Moriarty {\it et al.}

We may expect a proliferation of kinks at a stress of the order of
\begin{equation}
\tau_0 = \frac{E^{\rm kink}}{b L^{\rm kink} l_P}
\end{equation}
for which $\Delta G = 0$. For Mo, Xu and Moriarty
\cite{XuMoriarty1998} have computed $\tau_0$ to be of the order of
a few GPa. The activation free enthalpy for double kink nucleation
may thus be rewritten as
\begin{equation}
\Delta G^\pm(\tau) = E^{\rm kink} \left( 1 \mp
\frac{\tau}{\tau_0}\right), \quad 0 \leq \tau < \tau_0
\label{ActivationEnthalpy}
\end{equation}
A slightly more elaborate empirical formula that is widely used to
fit activation energies was proposed by Kocks {\it et al.}
\cite{KocksArgonAshby1975} and the corresponding parameters have
been determined for Ta by Tang {\it al.} \cite{TangKubinCanova1998},
but this enhancement will not be pursued here.

Transition state theory predicts that a dislocation segment effects
$\nu_D \text{e}^{-\beta G^+}$ and $\nu_D \text{e}^{-\beta G^-}$ jumps
per unit time in the positive and negative directions,
respectively. Here, $\beta = 1/k_B T$, $k_B$ is Boltzmann's constant,
$T$ is the absolute temperature, and $\nu_D$ is the attempt frequency
which may be identified with the Debye frequency to a first
approximation. Since the length of the individual jumps is $l_P$, the
mean velocity of the dislocations then follows as
\begin{equation}\label{Velocity}
v = 2 l_P \nu_D {\rm e}^{- \beta E^{\rm kink}} \sinh\left(\beta E^{\rm
kink} \frac{\tau}{\tau_0} \right)
\end{equation}
An application of Orowan's formula then gives the slip strain rate as
\begin{equation}\label{GammaDot}
\dot{\gamma} = \dot{\gamma}^{\rm kink}_0 {\rm e}^{- \beta E^{\rm
kink}} \sinh\left(\beta E^{\rm kink} \frac{\tau}{\tau_0} \right)
\end{equation}
where
\begin{equation}
\dot{\gamma}^{\rm kink}_0 = 2 b \rho l_P \nu_D
\end{equation}
is a reference strain rate, and $\rho$ is the dislocation density. In
writing \eqref{GammaDot}, we have taken into account the possibility
of thermally activated jumps in the reverse direction, with forward
and backward activation enthalpies $\Delta G^+$ and $\Delta G^-$,
respectively. For slip in the positive direction, $\dot{\gamma} \geq
0$, (\ref{GammaDot}) may be inverted to give:
\begin{equation}
\frac{\tau_P}{\tau_0} = \frac{1}{\beta E^{\rm
kink}} \asinh\left(\frac{\dot{\gamma}}{\dot{\gamma}^{\rm kink}_0}{\rm
e}^{\beta E^{\rm kink}}\right)
\label{YieldStressDependence}
\end{equation}
where $\tau_P$ may be regarded as a temperature and rate-dependent
\emph{effective} Peierls stress.

Figure 2 illustrates the dependence of the
effective Peierls stress on temperature and rate of deformation.
The Peierls stress decreases ostensibly linearly up to a critical
temperature $T_c$, beyond which it rapidly tends to zero. These
trends are in agreement with the experimental observations of
Wasserb\"ach \cite{Wasserbach1986} and Lachenmann and Schultz
\cite{LachenmannShultz1970}. The critical temperature $T_c$
increases with the strain rate. In particular, in this model the
effect of increasing (decreasing) the strain rate has an analogous
effect to decreasing (increasing) the temperature, and vice-versa,
as noted by Tang {\it et al.} \cite{TangDevincreKubin1999}. In the
regime of very high strain-rates ($\dot{\gamma} > 10^5 \:
s^{-1}$), effects such as electron and phonon drag become
important and control the velocity of dislocations
\cite{suzuki:1991, brailsford:1969}.

\section{Forest hardening}

In the forest-dislocation theory of hardening, the motion of
dislocations, which are the agents of plastic deformation in
crystals, is impeded by secondary --or `forest'-- dislocations
crossing the slip plane. As the moving and forest dislocations
intersect, they form jogs or junctions of varying strengths
\cite{baskes:1998, RodneyPhillips1999, PhillipsRodneyShenoy1999,
ShenoyKuktaPhillips2000, DannaBenoit1993, RheeZbibHirth1998,
HuangGhoniemDelaRubia1999, KubinDevincreTang1998,
ZbibDeLaRubiaRhee2000} which, provided the junction is
sufficiently short, may be idealized as point obstacles. Moving
dislocations are pinned down by the forest dislocations and
require a certain elevation of the applied resolved shear stress
in order to bow out and bypass the pinning obstacles. The net
effect of this mechanism is macroscopic hardening. Detailed
numerical simulations of a dislocation line propagating through
forest dislocations have been carried out by Foreman and Makin
\cite{foreman:1966, foreman:1967}, and by Kocks \cite{kocks:1966},
and more recently by Tang {\it et al.}
\cite{TangDevincreKubin1999}. Analytical treatments of the model
have been given by Kocks \cite{kocks:1966}, Ortiz and Popov
\cite{ortiz:1982}, and Cuiti\~no and Ortiz \cite{cuitino:1992} for
the case of infinitely strong obstacles. A phase-field model of
the forest hardening mechanism has been proposed by Ortiz
\cite{Ortiz1999} and by Cuiti\~no {\it et al.} \cite{Cuitino2000}.
Here we extend the statistical analysis of Cuiti\~no and Ortiz
\cite{cuitino:1992} to the case of several species of obstacles of
finite strength.

Because of the random nature of the interactions, the motion of
dislocations through a distribution of obstacles is best described
in statistical terms \cite{ortiz:1982}. We begin by treating the
case of infinitely strong obstacles. In this case, pairs of
obstacles pin down dislocation segments, which require a
certain threshold resolved shear stress $s$ in order to overcome
the obstacle pair. Since the distribution of point obstacles
within the slip plane is random, it follows that $s$ is itself a
random variable. We shall let $\tilde{f}_0^{\alpha}(s,t)$ denote
the probability density function of two-point barrier strengths on
slip system $\alpha$ at time $t$. The time dependence of
$\tilde{f}_0^{\alpha}(s,t)$ is a consequence of the variation in
forest dislocation density.

In order to determine the precise form of
$\tilde{f}_0^{\alpha}(s,t)$, we begin by noting that the Peierls
energy landscape of \bcc crystals strongly favors either screw or
edge segments \cite{ChangBulatovYip1999}. In addition kinks, or
points of change of direction of the dislocation line, carry a
non-negligible amount of energy.  It therefore follows that the
lowest-energy configuration of unstressed dislocation segments
spanning an obstacle pair is a step of the form shown in
Fig.~3.  Under these conditions, the bow-out
mechanism by which a dislocation segment bypasses an obstacle pair
may be expected to result in the configuration shown in
Fig.~3 (bold line).  If the edge-segment length
is $l_{e}$, a displacement $da_{e}$ of the dislocation requires a
supply of energy equal to $2 U^{\rm screw} da_{e} + b \tau_P^{\rm
edge} l_{e} da_{e}$ in order to overcome the Peierls resistance
$\tau_P^{\rm edge}$ and to extend the screw segments.  The
corresponding energy release is $b \tau l_{e} da_{e}$.  Similar
contributions result from a displacement $da_{s}$ of the
screw-segment of length $l_{s}$. Therefore, for the bow-out of the
dislocation to be energetically possible we must have
\begin{equation}\label{TauC0}
b \tau l_{e} da_{e} + b \tau l_{s} da_{s} \geq 2 U^{\rm screw}
da_{e} + b \tau_P^{\rm edge} l_{e} da_{e} + 2 U^{\rm edge}  da_{s}
+ b \tau_P^{\rm screw}l_{s} da_{s}
\end{equation}
As already noted, in \bcc crystals the core energy $U^{\rm screw}$ per
unit length of the screw segments is smaller than the core energy
$U^{\rm edge}$ per unit length of the edge segments (e.~g.,
\cite{WangStrachanCaginGoddard2000}). Conversely, the Peierls resistance
$\tau_P^{\rm screw}$ to the glide of screw segments is larger than the
Peierls resistance $\tau_P^{\rm edge}$ for edge segments (e.~g.,
\cite{DuesberyXu1998}). Retaining dominant terms only,
Eq.~\eqref{TauC0} simplifies to
\begin{equation}\label{TauC}
\tau \geq s = \tau_P^{\rm screw} + \frac{2 U^{\rm edge}}{b l_{s}}
\end{equation}
We may further identify the Peierls stress $\tau_P^{\rm screw}$
with (\ref{YieldStressDependence}), whence it follows that $s$
comprises a thermally-activated and rate-sensitive term $\tau_P$
and an athermal and rate-insensitive term $2 U^{\rm edge}/b l_{s}$.

We shall assume that the point obstacles are randomly distributed
over the slip plane with a mean density $n^\alpha$ of obstacles
per unit area. We shall also assume that the obstacle pairs
spanned by dislocation segments are nearest-neighbors in the
obstacle ensemble. Thus, if $r$ is the distance between the
obstacles in an obstacle pair, then the circle of radius $r$
centered at either obstacle contains no other obstacles. Under
these conditions, the probability density of $r$ is given by
\cite{kocks:1966, cuitino:1992}
\begin{equation}\label{FR}
\tilde{f}_0^\alpha(r,t) = 2 \pi n^\alpha r \exp( - \pi n^\alpha
r^2 )
\end{equation}
where the dependence of $\tilde{f}_0$ on time stems from the
time-dependence of the obstacle density $n^\alpha$.

In order to deduce the probability density of $l_{s}$, and by
extension of $s$, we note that, given a point obstacle located at the
origin, the probability of finding another obstacle in the element of
area $r dr d\theta$ is:
\begin{equation}\label{DP}
d\tilde{P}_0^\alpha(r,t) = n^\alpha r \exp( - \pi n^\alpha r^2 )
dr d\theta
\end{equation}
Changing variables to Cartesian coordinates, $x = r \cos\theta$,
$y = r \sin\theta$, (\ref{DP}) may be recast in the form:
\begin{equation}\label{DP2}
d\tilde{P}_0^\alpha = n^\alpha \exp[ - \pi n^\alpha (x^2 + y^2) ]
dx dy
\end{equation}
The frequency of obstacle pairs with screw-segment length $l_s$ is,
therefore:
\begin{equation}\label{FL}
\tilde{f}_0^\alpha(l_s,t) = 2 \int_{-\infty}^\infty n^\alpha \exp[ -
\pi n^\alpha (x^2 + l_s^2) ] dx = \sqrt{n^\alpha}  \exp( - \pi
n^\alpha l_s^2 )
\end{equation}
The corresponding probability of finding an obstacle pair of
strength $s$ follows from (\ref{FL}) and (\ref{TauC}) as:
\begin{equation}\label{FS}
\tilde{f}_0^\alpha(s,t) =  \frac{4 U^{\rm edge} \sqrt{n^\alpha}}{b
(s - \tau_P^\alpha)^2} \exp\left[ - \frac{ \pi n^\alpha}{b^2}
\left( \frac{ 2 U^{\rm edge}}{s - \tau_P^\alpha} \right)^2
\right], \quad s \geq \tau_P^\alpha
\end{equation}
and the associated distribution function is:
\begin{equation}\label{PS}
\tilde{P}_0^\alpha(s,t) = \int_{\tau_P}^s
\tilde{f}_0^\alpha(\xi,t) d\xi = \left[ 1 - {\rm erf} \left(
\frac{2 \sqrt{\pi n^\alpha} U^{\rm edge}}{b (s - \tau_P^\alpha)}
\right) \right]
\end{equation}
We note that the Peierls stress $\tau^\alpha_P$ depends on the
slip system $\alpha$ through its dependence on the slip-strain
rate $\dot{\gamma}^\alpha$.

It is interesting to note that the probability density
$\tilde{f}_0^\alpha(s,t)$ of obstacle-pair strengths just derived
for \bcc crystals differs markedly from those which are obtained
for \fcc crystals by a similar argument \cite{kocks:1966,
mughrabi:1975, grosskreutz:1975, cuitino:1992}, namely:
\begin{equation}\label{FSFCC}
\tilde{f}_0^\alpha(s,t) = \frac{2 \pi n^\alpha U^2}{b^2 s^3}
\exp\left( - \frac{n^\alpha \pi}{b^2} \frac{U^2}{s^2} \right)
\end{equation}
where $U$ is the dislocation core energy per unit length of
dislocation.  This difference owes to the different bow-out
configurations for the two crystal classes and the comparatively
larger values of the Peierls stress in \bcc crystals.  Thus, the
Peierls stress of \fcc crystals is generally quite small and, as
in the derivation of (\ref{FSFCC}) is often neglected entirely to
a first approximation.  For \bcc crystals, the effective Peierls
stress $\tau_P^\alpha$ decreases with increasing temperature and
decreasing rate of deformation, Eq.~(\ref{YieldStressDependence}),
and hence the behavior of \bcc crystals may be expected to be
closer to that of \fcc crystals under those conditions, as noted
by Tang {\it et al.} \cite{TangDevincreKubin1999}.  Conversely,
the hardening behavior of \bcc crystals may be expected to differ
sharply from that of \fcc crystals at low temperatures and high
rates of deformation.

The function $\tilde{f}_0^\alpha(s,t)$ just derived provides a
complete description of the distribution of the obstacle-pair
strengths when the point obstacles are of infinite strength and,
consequently, impenetrable to the dislocations.  Next we extend
the preceding analysis to the case of finite obstacle strengths.
Let $s^{\alpha\beta}$ be the strength of the jogs or junctions
formed by dislocations of systems $\alpha$ and $\beta$.  Now
consider an obstacle pair in which the weakest point obstacle
corresponds to a forest dislocation of type $\beta$ and,
therefore, has strength $s^{\alpha\beta}$. The probability that
the strength of the obstacle pair be $s$ is, therefore,
\begin{equation}
\tilde{f}^\alpha(s | s^{\alpha\beta},t) =
\dfrac{\tilde{f}^\alpha_0(s,t)}
{\tilde{P}^\alpha_0(s^{\alpha\beta},t)} \left[ 1 - H(s -
s^{\alpha\beta})\right] \label{FiniteStrengthProbability}
\end{equation}
where $H(s)$ is the Heaviside function.  The probability that an
obstacle on system $\alpha$ be of type $\beta \neq \alpha$ is
$n^{\alpha\beta}/n^\alpha$, where $n^{\alpha\beta}$ is the number
of obstacles of type $\beta$ per unit area of the slip plane
$\alpha$, and
\begin{equation}
n^\alpha(t) = \sum_{\beta\neq\alpha} n^{\alpha\beta}(t) \label{NA}
\end{equation}
is the total obstacle density on slip system $\alpha$.

Next we note that the probability that the weakest of the two
obstacles forming a obstacle pair be of type $\beta$ is:
\begin{equation}
p^{\alpha\beta} = \frac{n^{\alpha\beta}}{n^\alpha} \left(
\frac{n^{\alpha\beta}}{n^\alpha} + 2 \sum_{s^{\alpha\beta'}
> s^{\alpha\beta}}
\frac{n^{\alpha\beta'}}{n^\alpha} \right), \quad \beta \neq
\alpha, \quad \beta' \neq \alpha \label{PAB}
\end{equation}
It is readily verified that
\begin{equation}
\sum_{\beta\neq\alpha} p^{\alpha\beta} =
\left[\sum_{\beta\neq\alpha} \left(\frac{n^{\alpha\beta}}
{n^\alpha}\right)\right]^2 = 1 \label{Normalization}
\end{equation}
as required. Finally, the probability that an obstacle pair have
strength $s$ follows as
\begin{equation}
\tilde{f}^\alpha(s,t) = \sum_{\beta\neq\alpha} p^{\alpha\beta}(t)
\tilde{f}^\alpha(s | s^{\alpha\beta},t) \label{Finite}
\end{equation}
This probability distribution function jointly accounts for the
strength of the obstacle pairs due to line tension and to the
obstacle strength. Evidently, $\tilde{f}^\alpha(s,t)$ is supported
in the interval $[0,s_{\rm max}]$, where $s_{\rm max}$ is the
maximum obstacle strength. Making use of
(\ref{FiniteStrengthProbability}), the probability density
function \eqref{Finite} can be rewritten in the form
\begin{equation}
\tilde{f}^\alpha(s,t) = \sum_{\beta\neq\alpha}
p^{\alpha\beta}(t)\; \tilde{f}_0^\alpha(s,t) \;
\frac{1-H(s-s^{\alpha\beta})}
{\tilde{P}_0^\alpha(s^{\alpha\beta},t)}
\label{ProbaStrengthFinite}
\end{equation}
and the associated distribution function becomes
\begin{equation}
\tilde{P}^\alpha(s,t) = \sum_{\beta\neq\alpha} p^{\alpha\beta}(t)
\; \left\{ \tilde{P}_0^\alpha(s,t) \;
\frac{1-H(s-s^{\alpha\beta})}
{\tilde{P}_0^\alpha(s^{\alpha\beta},t)} +
H(s-s^{\alpha\beta})\right\}  \label{DistribStrengthFinite}
\end{equation}

We assume that $n^{\alpha\beta}$, the number of obstacles of type
$\beta$ per unit area of the slip plane $\alpha$, scales with the
dislocation density $\rho^\beta$ according to the relation:
\begin{equation}
n^{\alpha\beta} = a^{\alpha\beta} \, \rho^\beta \label{NABRho2}
\end{equation}
where the coupling constants $a^{\alpha\beta}$ are regarded here
as purely geometrical parameters.  A simple geometrical argument
based on counting intersections of randomly distributed lines with
a slip plane gives \cite{cuitino:1993}:
\begin{equation}
  n^{\alpha\beta} = \frac{2}{\pi}
    \sqrt{1-(\bm^\alpha\cdot\bm^\beta)^2} \, \rho^\beta
\label{NABRho}
\end{equation}
where $\bm^\alpha$ and $\bm^\beta$ are the unit normals to slip
planes of types $\alpha$ and $\beta$ and $\rho^\beta$ is the
dislocation line density per unit volume in slip system $\beta$. A
comparison between (\ref{NABRho2}) and (\ref{NABRho}) suggests
writing:
\begin{equation}
  a^{\alpha\beta} = a_0 \frac{2}{\pi}
    \sqrt{1-(\bm^\alpha\cdot\bm^\beta)^2}
\label{NABRho3}
\end{equation}
where $a_0 < 1$ is an `efficiency' factor which accounts for the
tendency of dislocations to tangle and form loops, which in turn
tends to lower the number of slip-plane crossings.

Let $\tau^\alpha(t)$ now be the resolved shear stress acting on
the slip system $\alpha$ at time $t$.  Assume for now that
$\tau^\alpha(t)$ increases monotonically from zero at $t=0$.
Evidently, for dislocations to be stable at time $t$ they must
face barriers of strengths $s$ in excess of $\tau^\alpha(t)$.  As
$\tau^\alpha(t)$ is increased to $\tau^\alpha(t) +
\dot\tau^\alpha(t)dt$, the dislocation segments held at barriers
of strengths in the range $\tau^\alpha(t)\leq
s\leq\tau^\alpha(t)+\dot\tau^\alpha(t)dt$ are dislodged and move
forward until they reach barriers of strength
$s\geq\tau^\alpha(t)+\dot\tau^\alpha(t)dt$. This motion of
dislocations results in an net increase in the plastic
deformation. The dislocation jumps between obstacles are assumed
to be instantaneous. This idealization is justified when the
duration of the flights is much smaller than the characteristic
time of variation of the loads.

As noted by Ortiz and Popov~\cite{ortiz:1982}, the information
needed to describe the dislocation motion is fully contained in
the probability density function $f^\alpha(s,t)$, which represents
the fraction of dislocation length facing obstacle pairs of
strength $s$ at time $t$. The function $f^\alpha(s,t)$ evolves in
time due to the process of redistribution of dislocation line
described above. Initially, though, the dislocations may be
assumed to be randomly distributed over their slip plane, and
$f^\alpha(s,0) = \tilde{f}^\alpha(s,0)$. At later times,
$f^\alpha(s,t)$ must vanish identically for $0\leq
s<\tau^\alpha(t)$ in the rate independent limit.

A kinetic equation for the evolution of $f^\alpha(s,t)$ was derived by
Ortiz and Popov~\cite{ortiz:1982} using standard tools of
non-equilibrium statistical mechanics. Cuiti\~no and Ortiz
\cite{cuitino:1992} were able to obtain an analytical solution to
this kinetic equation for the case of monotonic loading and an
arbitrary time variation of the density of point
obstacles. Remarkably, the solution takes the simple closed form
\begin{equation}
f^\alpha(s,t) = \frac{\tilde{f}^\alpha(s,t)}{1-
\tilde{P}^\alpha\left(\tau^\alpha(t),t\right)}\;
H\left(s-\tau^\alpha(t)\right) \label{DislocationDistribution}
\end{equation}
Solution \eqref{DislocationDistribution} implies that, under the
conditions of the analysis, the probability density $f^\alpha(s,t)$
remains proportional to $\tilde{f}^\alpha(s,t)$ over the current
admissible range $[\tau^\alpha(t),s_{\text{max}}]$.

Let $\rho^\alpha(t)$ denote the current dislocation length per unit
volume for the slip system $\alpha$. The dislocation density released
during an increment of the resolved shear stress from $\tau^\alpha(t)$
to $\tau^\alpha(t) + \dot\tau^\alpha(t)dt$ gives rise to an
incremental plastic strain \cite{kocks:1966, Teodosiu1970}
\begin{equation}
d\gamma^\alpha(t) = b\rho^\alpha(t)
f\left(\tau^\alpha(t),t)\right) \dot{\tau}^\alpha(t) dt
\bar{N}(t)\bar{l}(t) \label{StrainIncrement}
\end{equation}
where $\bar{l}(t)$ is the average distance between obstacles, and
$\bar{N}(t)$ is the average number of jumps the dislocation
segments make before attaining stable positions.

To a good approximation, $\bar{l}(t)$ can be identified with the
average distance between point obstacles.  Assuming that the
obstacles are randomly distributed over the slip plane with
density $n^\alpha(t)$, a straightforward derivation from
\eqref{FR} gives \cite{cuitino:1992}
\begin{equation}\label{MeanFreePath}
\bar{l}(t) = \langle l\rangle(t) = \frac{1}{2\sqrt{n^\alpha(t)}}
\end{equation}
Next we compute the average number of jumps $\bar{N}(t)$.  Evidently,
the probability that an unstable segment becomes arrested after the
first jump is equal to the probability that the first barrier
encountered is of a strength $s\geq \tau^\alpha(t)$. This probability
is $1-\tilde{P}^\alpha\left(\tau^\alpha(t),t\right)$. The probability
that the segment goes beyond the first barrier is
$\tilde{P}^\alpha\left(\tau^\alpha(t),t\right)$.  Likewise, the
probability that a segment gets arrested at the second barrier
encountered is $\tilde{P}^\alpha\left(\tau^\alpha(t),t\right)
[1-\tilde{P}^\alpha\left(\tau^\alpha(t),t\right)]$, and the
probability that it goes beyond is
$\tilde{P}^\alpha{}^2\left(\tau^\alpha(t),t\right)$, and so on.
Hence, the average number of jumps between barriers taken by an
unstable segment is
\begin{equation}
\begin{split}
  \bar{N}(t) &= [1-\tilde{P}^\alpha\left(\tau^\alpha(t),t\right)]
     + 2\tilde{P}^\alpha\left(\tau^\alpha(t),t\right)
              [1-\tilde{P}^\alpha\left(\tau^\alpha(t),t\right)] \\
    &+ 3\tilde{P}^\alpha{}^2\left(\tau^\alpha(t),t\right)
              [1-\tilde{P}^\alpha\left(\tau^\alpha(t),t\right)]
     + \ldots
   = \frac{1}{1-\tilde{P}^\alpha\left(\tau^\alpha(t),t\right)}
\end{split}
\label{Jumps}
\end{equation}
Interestingly, if $\tau^\alpha =0$, any moving dislocation segment is
sure to be arrested at the first obstacle it encounters, and $\bar{N}
= 1$, in agreement with \eqref{Jumps}. Likewise, if
$\tau^\alpha(t)>s_{\rm max}$, then the segment never reaches a
stable barrier and $\bar{N}\rightarrow\infty$, as predicted by
\eqref{Jumps}.

Substituting \eqref{Jumps} and \eqref{DislocationDistribution} with
$s=\tau^\alpha(t)$ into \eqref{StrainIncrement} we finally obtain
\begin{equation}
\dot{\gamma}^\alpha(t) = \gamma_{\text{c}}^\alpha(t) \;
\frac{\tilde{f}^\alpha(\tau^\alpha(t),t)}{ [ 1 -
\tilde{P}^\alpha(\tau^\alpha(t),t)]^2} \; \dot{\tau}^\alpha(t)
\label{SlipStrain}
\end{equation}
where the characteristic plastic strain $\gamma_{\text{c}}^\alpha(t)$
is defined as
\begin{equation}
\gamma_{\text{c}}^\alpha(t) = \frac{b \rho^\alpha(t)}{2
\sqrt{n^\alpha(t)}} \label{GammaC}
\end{equation}
Equation (\ref{SlipStrain}) defines a relation of the form
\begin{equation}
\dot{\gamma}^\alpha(t) = \frac{\dot{\tau}^\alpha(t)}{h^\alpha(t)}
\end{equation}
where
\begin{equation}\label{H}
h^\alpha(t) = \frac{1}{\gamma_{\text{c}}^\alpha(t)} \; \frac{ [ 1
- \tilde{P}^\alpha(\tau^\alpha(t),t)]^2}
{\tilde{f}^\alpha(\tau^\alpha(t),t)}
\end{equation}
is the hardening modulus of the slip system $\alpha$. By way of
example, in the particular case of obstacles of uniform strength,
$s^{\alpha\beta} = s_{\text{max}}$, the hardening modulus takes
the form:
\begin{multline}
h^\alpha(t) = h_{c}^{\alpha}(t)
\left(\frac{\tau^{\alpha}(t)-\tau_{P}^{\alpha}}{\tau_{c}^{\alpha}}\right)^{2}
\frac{1}{\sqrt{\pi}} \exp\left[\left(
\frac{\tau_c^{\alpha}}{\tau^{\alpha}(t)-\tau_{P}^{\alpha}}\right)^{2}\right]\\
\frac{[\tilde{P}_0^{\alpha}(s_{\rm max},t)
-\tilde{P}_0^{\alpha}(\tau^{\alpha},t)]^{2}}{\tilde{P}_0^{\alpha}(s_{\rm
max},t)} (1-H(\tau^{\alpha}-s_{\rm max}))
\label{SelfHardeningMono}
\end{multline}
where
\begin{equation}
\tau_{\text{c}}^\alpha(t) = 2\sqrt{\pi n^{\alpha}} \frac{U^{\rm
edge}}{b}
  \quad\text{and}\quad
h_{\text{c}}^\alpha(t) =
\frac{\tau_{\text{c}}^\alpha(t)}{\gamma_{\text{c}}^\alpha(t)}
\end{equation}
are a characteristic shear stress and plastic modulus,
respectively. Equation~(\ref{SelfHardeningMono}) predicts an
initial infinite hardening modulus at $\tau=\tau_P$. The hardening
modulus subsequently decreases monotonically to zero as $\tau$
approaches $s_{\text{max}}$.

As shown in \cite{cuitino:1992}, the above relations can be
extended simply to account for elastic unloading, provided that
$\tau^\alpha(t)$ does not change sign at any time during the
loading history. This extension consists in defining the current
flow stress $g^\alpha(t)$ of slip system $\alpha$ as the maximum
previously attained value of $\tau^\alpha(t)$. The hardening
relations are then rewritten in the form
\begin{equation}
\dot{\gamma}^\alpha(t) = \frac{\dot{g}^\alpha(t)}{h^\alpha(t)}
\label{StrainRate}
\end{equation}
and
\begin{equation}
h^\alpha(t) = \frac{1}{\gamma_{\text{c}}^\alpha(t)} \; \frac{ [ 1
- \tilde{P}^\alpha(g^\alpha(t),t)]^2}
{\tilde{f}^\alpha(g^\alpha(t),t)}
\end{equation}
It should be noted that, since $\dot{g}^\alpha(t) = 0$ whenever
either $\tau^\alpha(t) < g^\alpha(t)$ or $\dot{\tau}^\alpha(t)
\leq 0$, relation~(\ref{StrainRate}) implicitly accounts for
elastic unloading. To account for the initial distribution of
dislocations and obstacles, we take $g^\alpha(t)=g_0$.

\section{Dislocation intersections}

In this section we proceed to estimate the obstacle strengths
which enter in the preceding analysis of forest hardening. The
interaction between primary and secondary dislocations may result
in a variety of reaction products, including jogs and junctions
\cite{DannaBenoit1993, RheeZbibHirth1998, baskes:1998,
HuangGhoniemDelaRubia1999, TangDevincreKubin1999,
RodneyPhillips1999, PhillipsRodneyShenoy1999,
ZbibDeLaRubiaRhee2000, ShenoyKuktaPhillips2000}. Experimental
estimates of junction strengths have been given by Franciosi and
Zaoui \cite{franciosi:1982} for the twelve slip systems belonging
to the family of $\{111\}$ planes and $[110]$ directions in \fcc
crystals, and by Franciosi \cite{franciosi:1983} for the
twenty-four systems of types $\{211\}$ $[111]$ and $\{110\}$
$[111]$ in \bcc crystals.  The strength of some of these
interactions has recently been computed using atomistic and
continuum models \cite{baskes:1998, RodneyPhillips1999,
PhillipsRodneyShenoy1999, ShenoyKuktaPhillips2000}. Tang {\it et
al.} have numerically estimated the average strength of
dislocation junctions for Nb and Ta crystals
\cite{TangDevincreKubin1999}.

For purposes of the present theory, we specifically concern
ourselves with short-range interactions between dislocations which
can be idealized as point defects. For simplicity, we consider the
case in which each intersecting dislocation acquires a jog. The
energy of a pair of crossing dislocations is schematically shown
in Fig.~4 as a function of some convenient
reaction coordinate, such as the distance between the
dislocations. The interaction may be repulsive, resulting in an
energy barrier, or attractive, resulting in a binding energy,
Fig.~4. In the spirit of an equilibrium theory,
here we consider only the final reaction product, corresponding to
a pair of jogged dislocations at infinite distance from each
other, and neglect the intermediate states along the reaction
path. In addition, we deduce the strength of the obstacles
directly from the energy supply required to attain the final
state, i.~e. the jog-formation energy. Despite the sweeping nature
of these assumptions, the predicted saturation strengths in
multiple slip are in good agreement with experiment (cf
Section~\ref{ComparisonExperiment}), which lends some empirical
support to the theory.

We estimate the jog formation energy as follows. Based on energy
and mobility considerations already discussed, we may expect the
preponderance of forest dislocations to be of screw character, and
the mobile dislocation segments to be predominantly of edge
character. We therefore restrict our analysis to intersections
between screw and edge segments. The geometry of the crossing
process is schematically shown in Fig.~5. Each
dislocation acquires a jog equal to the Burgers vector of the
remaining dislocation. The energy expended in the formation of the
jogs may be estimated as
\begin{equation}\label{EJogs}
E^{\rm jog}_{\alpha\beta} \sim |\bb^{\beta}| \left[
U^{\alpha\beta} - U^{\alpha\alpha} \cos\theta^{\alpha\beta}
\right] + |\bb^{\alpha}| \left[ U^{\beta\alpha} - U^{\beta\beta}
\cos\theta^{\beta\alpha} \right]
\end{equation}
Here, $\alpha$ designates the slip system of the moving edge
segment, $\beta$ the slip system of the forest screw dislocation,
$U^{\alpha\alpha} = U^{\rm edge}$ is the energy per unit length of
an edge segment in the slip system $\alpha$, $U^{\beta\beta} =
U^{\rm screw}$ is the energy per unit length of a screw segment in
slip system $\beta$, $U^{\alpha\beta}$ is the energy per unit
length of a segment in slip system $\alpha$ aligned with
$\bb^\beta$, $U^{\beta\alpha}$ is the energy per unit length of a
segment in slip system $\beta$ aligned with $\bb^\alpha$,
$\theta^{\alpha\beta}$ is the angle between the edge direction in
system $\alpha$ and $\bb^\beta$, and $\theta^{\beta\alpha}$ is the
angle between the screw direction in system $\beta$ and $\bb^\alpha$.

We additionally assume that the screw orientation defines a cusp
in the dependence of the dislocation line energy on segment
orientation, and that small deviations of a segment from a pure
screw character raise its energy to a level comparable to $U^{\rm
edge}$. This gives the energy estimate
\begin{equation}
U^{\alpha\beta} = U^{\beta\alpha} =
\begin{cases}
U^{\rm screw} & \text{if } \bb^\alpha = \bb^\beta \\
U^{\rm edge} & \text{otherwise}
\end{cases}
\label{ener}
\end{equation}
Inserting these energies into (\ref{EJogs}) gives
\begin{equation}
E^{jogs}_{\alpha\beta} \sim
\begin{cases}
b U^{\rm screw} \left[ 1 - r \cos\theta^{\alpha\beta} \right] &
\text{if } \bb^\alpha = \bb^\beta \\
b U^{\rm screw} \left[ 2 r - \cos(\theta^{\beta\alpha}) - r
\cos\theta^{\alpha\beta} \right] & \text{otherwise}
\end{cases}
\label{energies}
\end{equation}
where $r = U^{\rm edge}/U^{\rm screw}$ is the ratio of screw to
edge dislocation line energies. For Ta, recent atomistic
calculations \cite{WangStrachanCaginGoddard2000} give a value 
of $r = 1.77$. The
resulting jog formation energies for the complete collection of
pairs of $\{211\}$ and $\{110\}$ dislocations are tabulated in
Table~\ref{tab:matprop3}.

\bigskip

\begin{table}
\centering
{\footnotesize{
\begin{tabular}
{ | l | l@{} l@{} l@{} l@{} l@{} l@{} l@{} l@{} l@{} l@{} l@{} l@{}
  l@{} l@{} l@{} l@{} l@{} l@{} l@{} l@{} l@{} l@{} l@{} l@{} |}
\hline
\space & {\bfseries A2\space} & {\bfseries A2'} & {\bfseries
A3\space} & {\bfseries A3'} &
     {\bfseries A6\space} & {\bfseries A6'} & {\bfseries B2\space} &
{\bfseries B2''} &
     {\bfseries B4\space} & {\bfseries B4'} & {\bfseries B5\space} &
{\bfseries B5'} &
     {\bfseries C1\space} & {\bfseries C1'} & {\bfseries C3\space} &
{\bfseries C3''} &
     {\bfseries C5\space} & {\bfseries C5''} & {\bfseries D1\space} &
{\bfseries D1''} &
     {\bfseries D4\space} & {\bfseries D4''} & {\bfseries D6\space} &
  {\bfseries D6''} \\ \hline
{\bfseries A2}   &  ---- &  1.0 &  1.0 &  1.0 &  1.0 &  1.0 &  1.5 &
1.5 &  1.5 &  1.5 &  1.5 &  1.5 &  2.4 &  2.4 &  2.4 &  2.4 &  2.4 &
2.4 &  2.4 &  2.4 &  2.4 &  2.4 &  2.4 &  2.4 \\
{\bfseries A2'}  &  1.0 &  ---- &  1.0 &  1.0 &  1.0 &  1.0 &  3.2 &
3.2 &  3.2 &  3.2 &  3.2 &  3.2 &  1.8 &  1.8 &  1.8 &  1.8 &  1.8 &
1.8 &  1.8 &  1.8 &  1.8 &  1.8 &  1.8 &  1.8 \\
{\bfseries A3} &  1.0 &  1.0 &  ---- &  1.0 &  1.0 &  1.0 &  2.4 &
2.4 &  2.4 &  2.4 &  2.4 &  2.4 &  1.5 &  1.5 &  1.5 &  1.5 &  1.5 &
1.5 &  2.4 &  2.4 &  2.4 &  2.4 &  2.4 &  2.4 \\
{\bfseries A3'}  &  1.0 &  1.0 &  1.0 &  ---- &  1.0 &  1.0 &  1.8 &
1.8 &  1.8 &  1.8 &  1.8 &  1.8 &  3.2 &  3.2 &  3.2 &  3.2 &  3.2 &
3.2 &  1.8 &  1.8 &  1.8 &  1.8 &  1.8 &  1.8 \\
{\bfseries A6}  &  1.0 &  1.0 &  1.0 &  1.0 &  ---- &  1.0 &  2.4 &
2.4 &  2.4 &  2.4 &  2.4 &  2.4 &  2.4 &  2.4 &  2.4 &  2.4 &  2.4 &
2.4 &  1.5 &  1.5 &  1.5 &  1.5 &  1.5 &  1.5 \\
{\bfseries A6'} &  1.0 &  1.0 &  1.0 &  1.0 &  1.0 & ---- &  1.8 &
1.8 &  1.8 &  1.8 &  1.8 &  1.8 &  1.8 &  1.8 &  1.8 &  1.8 &  1.8 &
1.8 &  3.2 &  3.2 &  3.2 &  3.2 &  3.2 &  3.2 \\
{\bfseries B2} &  1.5 &  1.5 &  1.5 &  1.5 &  1.5 &  1.5 &  ---- &
1.0 &  1.0 &  1.0 &  1.0 &  1.0 &  2.4 &  2.4 &  2.4 &  2.4 &  2.4 &
2.4 &  2.4 &  2.4 &  2.4 &  2.4 &  2.4 &  2.4 \\
{\bfseries B2''} &  3.2 &  3.2 &  3.2 &  3.2 &  3.2 &  3.2 &  1.0 &
---- &  1.0 &  1.0 &  1.0 &  1.0 &  1.8 &  1.8 &  1.8 &  1.8 &  1.8 &
1.8 &  1.8 &  1.8 &  1.8 &  1.8 &  1.8 &  1.8 \\
{\bfseries B4}  &  2.4 &  2.4 &  2.4 &  2.4 &  2.4 &  2.4 &  1.0 &
1.0 &  ---- &  1.0 &  1.0 &  1.0 &  2.4 &  2.4 &  2.4 &  2.4 &  2.4 &
2.4 &  1.5 &  1.5 &  1.5 &  1.5 &  1.5 &  1.5 \\
{\bfseries B4'} &  1.8 &  1.8 &  1.8 &  1.8 &  1.8 &  1.8 &  1.0 &
1.0 &  1.0 &  ---- &  1.0 &  1.0 &  1.8 &  1.8 &  1.8 &  1.8 &  1.8 &
1.8 &  3.2 &  3.2 &  3.2 &  3.2 &  3.2 &  3.2 \\
{\bfseries B5}  &  2.4 &  2.4 &  2.4 &  2.4 &  2.4 &  2.4 &  1.0 &
1.0 &  1.0 &  1.0 &  ---- &  1.0 &  1.5 &  1.5 &  1.5 &  1.5 &  1.5 &
1.5 &  2.4 &  2.4 &  2.4 &  2.4 &  2.4 &  2.4 \\
{\bfseries B5'}  &  1.8 &  1.8 &  1.8 &  1.8 &  1.8 &  1.8 &  1.0 &
1.0 &  1.0 &  1.0 &  1.0 &  ---- &  3.2 &  3.2 &  3.2 &  3.2 &  3.2 &
3.2 &  1.8 &  1.8 &  1.8 &  1.8 &  1.8 &  1.8 \\
{\bfseries C1}  &  1.8 &  1.8 &  1.8 &  1.8 &  1.8 &  1.8 &  1.8 &
1.8 &  1.8 &  1.8 &  1.8 &  1.8 &  ---- &  1.0 &  1.0 &  1.0 &  1.0 &
1.0 &  3.2 &  3.2 &  3.2 &  3.2 &  3.2 &  3.2 \\
{\bfseries C1'} &  1.8 &  1.8 &  1.8 &  1.8 &  1.8 &  1.8 &  1.8 &
1.8 &  1.8 &  1.8 &  1.8 &  1.8 &  1.0 &  ---- &  1.0 &  1.0 &  1.0 &
1.0 &  3.2 &  3.2 &  3.2 &  3.2 &  3.2 &  3.2 \\
{\bfseries C3}  &  1.5 &  1.5 &  1.5 &  1.5 &  1.5 &  1.5 &  2.4 &
2.4 &  2.4 &  2.4 &  2.4 &  2.4 &  1.0 &  1.0 &  ---- &  1.0 &  1.0 &
1.0 &  2.4 &  2.4 &  2.4 &  2.4 &  2.4 &  2.4 \\
{\bfseries C3''}  &  3.2 &  3.2 &  3.2 &  3.2 &  3.2 &  3.2 &  1.8 &
1.8 &  1.8 &  1.8 &  1.8 &  1.8 &  1.0 &  1.0 &  1.0 &  ---- &  1.0 &
1.0 &  1.8 &  1.8 &  1.8 &  1.8 &  1.8 &  1.8 \\
{\bfseries C5} &  2.4 &  2.4 &  2.4 &  2.4 &  2.4 &  2.4 &  1.5 &
1.5 &  1.5 &  1.5 &  1.5 &  1.5 &  1.0 &  1.0 &  1.0 &  1.0 &  ---- &
1.0 &  2.4 &  2.4 &  2.4 &  2.4 &  2.4 &  2.4 \\
{\bfseries C5''}  &  1.8 &  1.8 &  1.8 &  1.8 &  1.8 &  1.8 &  3.2 &
3.2 &  3.2 &  3.2 &  3.2 &  3.2 &  1.0 &  1.0 &  1.0 &  1.0 &  1.0 &
---- &  1.8 &  1.8 &  1.8 &  1.8 &  1.8 &  1.8 \\
{\bfseries D1} &  1.8 &  1.8 &  1.8 &  1.8 &  1.8 &  1.8 &  1.8 &
1.8 &  1.8 &  1.8 &  1.8 &  1.8 &  3.2 &  3.2 &  3.2 &  3.2 &  3.2 &
3.2 &  ---- &  1.0 &  1.0 &  1.0 &  1.0 &  1.0 \\
{\bfseries D1''} &  1.8 &  1.8 &  1.8 &  1.8 &  1.8 &  1.8 &  1.8 &
1.8 &  1.8 &  1.8 &  1.8 &  1.8 &  3.2 &  3.2 &  3.2 &  3.2 &  3.2 &
3.2 &  1.0 &  ---- &  1.0 &  1.0 &  1.0 &  1.0 \\
{\bfseries D4}  &  2.4 &  2.4 &  2.4 &  2.4 &  2.4 &  2.4 &  1.5 &
1.5 &  1.5 &  1.5 &  1.5 &  1.5 &  2.4 &  2.4 &  2.4 &  2.4 &  2.4 &
2.4 &  1.0 &  1.0 &  ---- &  1.0 &  1.0 &  1.0 \\
{\bfseries D4''} &  1.8 &  1.8 &  1.8 &  1.8 &  1.8 &  1.8 &  3.2 &
3.2 &  3.2 &  3.2 &  3.2 &  3.2 &  1.8 &  1.8 &  1.8 &  1.8 &  1.8 &
1.8 &  1.0 &  1.0 &  1.0 &  ---- &  1.0 &  1.0 \\
{\bfseries D6} &  1.5 &  1.5 &  1.5 &  1.5 &  1.5 &  1.5 &  2.4 &
2.4 &  2.4 &  2.4 &  2.4 &  2.4 &  2.4 &  2.4 &  2.4 &  2.4 &  2.4 &
2.4 &  1.0 &  1.0 &  1.0 &  1.0 &  ---- &  1.0 \\
{\bfseries D6''} &  3.2 &  3.2 &  3.2 &  3.2 &  3.2 &  3.2 &  1.8 &
1.8 &  1.8 &  1.8 &  1.8 &  1.8 &  1.8 &  1.8 &  1.8 &  1.8 &  1.8 &
1.8 &  1.0 &  1.0 &  1.0 &  1.0 &  1.0 &  ---- \\ \hline

\end{tabular}}}

\caption{Normalized jog-formation energies resulting from
crossings of \bcc dislocations.} \label{tab:matprop3}
\end{table}
\bigskip

The net effect of jog formation on hardening may be ascertained as
follows. Consider the case in which a primary system $\alpha$
contains forest obstacles of a single species, corresponding to
secondary dislocations in the slip system $\beta$. Let
$n^{\alpha\beta}$ be the density of forest obstacles per unit area
of the primary plane $\alpha$. As a primary dislocation sweeps
through an area $A$, the energy expended in forming jogs with
forest dislocations of the $\beta$-type is $E^{\rm
jog}_{\alpha\beta} n^{\alpha\beta} A$. On the other hand, the
potential energy released as a result of the motion of the
dislocation follows from the Peach-Koehler formula as $b
\tau^\alpha A$, where $\tau^\alpha$ is the resolved shear stress
acting on the primary system $\alpha$. Hence, the forest obstacles
of type $\beta$ become `transparent' to the motion of primary
dislocations when $b \tau^\alpha A \geq E^{\rm jog}_{\alpha\beta}
n^{\alpha\beta} A$, or
\begin{equation}\label{JogTauC}
\tau^\alpha \ge 
\frac{1}{b} E^{\rm jog}_{\alpha\beta} n^{\alpha\beta}
\end{equation}

Since the jog energies scale with the elastic moduli, they may be
expected to reduce to zero at the melting temperature, which to
some extent accounts for the observed thermal softening. In
addition, the jog energies are small enough for thermal activation
to be operative at the level of individual obstacles. A derivation
entirely analogous to that leading to
Eq.~(\ref{YieldStressDependence}) yields, in this case,
\begin{equation}
\frac{s^{\alpha\beta}}{s^{\alpha\beta}_0} = \frac{1}{\beta E^{\rm
jog}_{\alpha\beta}} \asinh \left(
\frac{\dot{\gamma}^\alpha}{\dot{\gamma}^\alpha_0} {\rm e}^{\beta
E^{\rm jog}_{\alpha\beta}} \right)
\label{ObstacleStrengthDependence}
\end{equation}
where
\begin{equation}
s_0^{\alpha\beta} = \frac{E_{\alpha\beta}^{\rm jog}}{b
\bar{l}^{\alpha} L^{\rm junct}}
\end{equation}
and
\begin{equation}
\dot{\gamma}^\alpha_0 = 2 \rho^\alpha b \bar{l}^\alpha \nu_{D}
\end{equation}
The lengths $\bar{l}^\alpha$ and $L^{\rm junct}$ describe the geometry
of the junction, as illustrated in Fig.~6, and are
of the order of a few Burgers vectors.

It follows from these developments that, at low temperatures, the
energies collected in Table~\ref{tab:matprop3} provide a measure
of the corresponding obstacle strengths. We may also recall that
the saturation flow stress of a slip system is directly related to
the strengths of the obstacles. Consequently, information about
obstacle strengths may be inferred from saturation flow stresses
in crystals deformed in double slip. Measurements of this type
have been performed by Franciosi for $\alpha$-Fe
\cite{franciosi:1983}. The general trends exhibited by
Table~\ref{tab:matprop3} are consistent with Franciosi's data, as
well as with the recent analytical model of Lee {\it et al.}
\cite{LeeSubhashRavichandran1999}. In addition, the saturation
flow stresses resulting from the model developed above are in good
quantitative agreement with those measured experimentally, cf
Section~\ref{ComparisonExperiment}, which lends empirical support
to jog-formation as a plausible mechanism underlying short-range
obstacle strength and saturation in \bcc crystals.

\section{Dislocation evolution}

The density of forest obstacles depends directly on the
dislocation densities in all slip systems of the crystal.
Therefore, in order to close the model we require a equation of
evolution for the dislocation densities. Processes resulting in
changes in dislocation density include production by fixed
sources, such as Frank-Read sources, breeding by double cross slip
and pair annihilation (see \cite{kuhlmann:1989} for a review; see
also \cite{johnston:1959, johnston:1960, gillis:1965,
essmann:1973, Lagerlof1993, Dybiec1995}). Although the operation of 
fixed Frank-Read sources is quickly eclipsed by production
due to cross slip at finite temperatures, it is an important 
mechanisms at low temperatures. The double cross-slip, fixed
Frank-Read sources and pair annihilation mechanisms are next 
considered in turn.

\subsection{Breeding by double cross-slip}

The importance of breeding by double cross-slip as a dislocation
generation mechanism in crystals was emphasized by Johnston and
Gilman \cite{johnston:1959, johnston:1960}. In this mechanism, the
screw section of a moving dislocation migrates to a parallel plane
by double cross-slip, thus creating a pair of sessile segments
which pin down the dislocation and act in a manner similar to a
Frank-Read source.

The rate at which dislocation length is produced by this mechanism
can be estimated as follows. Let $N_{\rm CS}^\alpha$ denote the number of
dynamic sources per unit volume induced by double cross-slip on
slip system $\alpha$. Additionally, let $L^{\rm sat}$ denote the
dislocation length emitted by each source prior to saturation. The
rate of dislocation production per unit volume is, therefore:
\begin{equation}\label{DCG:RhoDot}
\dot{\rho}_{\rm CS}^\alpha = L^{\rm sat} \dot{N}_{\rm CS}^\alpha
\end{equation}
Assuming that cross-slip is thermally activated (see, e.~g.,
\cite{RasmussenJacobsenLeffers1997, RaoParthasarathyWoodward1999}
for recent calculations of pathways and energy barriers in
copper), the rate $\dot{N}_{\rm CS}^\alpha$ at which dynamic sources are
generated may in turn be computed as follows. Let $\bar{L}$ be the
mean-free path between cross-slip events, $L^{\rm cross}$ the
length of the screw segment effecting the double cross glide, and
$E^{\rm cross}$ the energy barrier for cross slip. The frequency
of cross glide attempts per unit volume is, therefore,
$\rho^\alpha v^\alpha/(\bar{L} L^{\rm cross})$, where $v^\alpha$
is the mean dislocation velocity. According to transition-state
theory, of these attempts a fraction ${\rm e}^{-\beta E^{\rm
cross}}$ is successful. This gives:
\begin{equation}\label{DCG:NDot}
\dot{N}_{\rm CS}^\alpha = \frac{\rho^\alpha v^\alpha}{\bar{L} L^{\rm
cross}} {\rm e}^{-\beta E^{\rm cross}}
\end{equation}
Indeed, theoretical \cite{li:1961} and experimental
\cite{johnston:1959, johnston:1960} investigations suggest that
the breeding rate due to cross-glide is proportional to the mean
dislocation speed. Using Orowan's formula, (\ref{DCG:NDot}) may be
rewritten as
\begin{equation}\label{DCG:NDot2}
\dot{N}_{\rm CS}^\alpha = \frac{\dot{\gamma}^\alpha}{b \bar{L} L^{\rm
cross}} {\rm e}^{-\beta E^{\rm cross}}
\end{equation}
Inserting this relation into (\ref{DCG:RhoDot}) gives
\begin{equation}\label{DCG:RhoDot2}
\dot{\rho}_{\rm CS}^\alpha = \frac{L^{\rm sat} \dot{\gamma}^\alpha }{b
\bar{L} L^{\rm cross}} {\rm e}^{-\beta E^{\rm cross}}
\end{equation}
Following Essmann and Rapp \cite{essmann:1973}, we shall
additionally assume that the mean free path is inversely
proportional to the dislocation density, which gives:
\begin{equation}\label{DCG:RhoDot3}
b \dot{\rho}_{\rm CS}^\alpha = \lambda_{\rm CS} \sqrt{\rho^\alpha}
\dot{\gamma}^\alpha
\end{equation}
where we have written
\begin{equation}\label{DCG:Lambda}
\lambda_{\rm CS} = \frac{L^{\rm sat}}{L^{\rm cross}} {\rm e}^{-\beta E^{\rm
cross}}
\end{equation}
An improvement on this model is to account for the resolved shear
stress acting on the cross-slip plane in the computation of the
activation energy, resulting in a so-called non-Schmid effect, but
this possibility will not be pursued here in the interest of
simplicity.

The double cross-slip mechanism is geometrically similar to the double
kink formation alluded to earlier. By virtue of this similarity,
we may expect that $L^{\rm cross} \approx L^{\rm kink}$ and
$E^{\rm cross} \approx E^{\rm kink}$ to a first approximation.

\subsection{Multiplication by fixed Frank-Read sources}

An infinite amount of dislocation multiplication can be sustained 
by the fixed Frank-Read sources as long as newly formed loops can 
expand and move away from the pinning points. The precise details of 
this mechanism, which was  independently proposed by 
Frank and Read\cite{FrankRead1950} in the early 50, are well documented
and can be consulted elsewhere\cite{hirth:1968}. In here, we limit our analysis
to obtained an estimate for the dislocation multiplication based on 
this mechanism. Considering that the obstacles in the slip plane
$\alpha$ serve as pinning points to operate
the Frank-Read sources, the increase of the dislocation population is
given by
\begin{equation}\label{DCG:RhoFRDot}
\dot{\rho}_{\rm FR}^\alpha = L_{\rm FR} \dot{N}_{\rm FR}^\alpha
\end{equation}
where ${N^{\alpha}_{\rm FR}}$ is the number of active sources 
per unit volume and $L^{\alpha}_{\rm FR}$ is the length of the emitted
segment by the source and $n^{\alpha}$ is the obstacle density given by
Eq.~(\ref{NA}). The effective rate of activation of Frank-Read sources
in a given system $\alpha$ can be estimated by the number of
intersections of the gliding loops with obstacles in that system
weighted by the efficiency of the source. In order to account
for the possibility of the newly formed loops to glide away from the
obstacles, we introduce the factor $\sqrt{\rho^\alpha/n^\alpha}$ which
is the ratio between mean distance between the obstacles and the mean
dislocation spacing. Then, 
\begin{equation}
\dot{N}^{\alpha}_{\rm FR} = 
\sqrt{\frac{\rho^\alpha}{n^\alpha}} n^{\alpha} v^{\alpha} \rho^{\alpha}.
\end{equation}
Introducing Orowan's formula and assuming that
the length of dislocation segments is proportional to the mean distance
between obstacles, the multiplication rate due to fixed Frank-Read
sources can be written as
\begin{equation}
b \dot{\rho}_{\rm FR}^\alpha = \lambda_{\rm FR} \sqrt{\rho^{\alpha}} \dot{\gamma}^{\alpha}
\label{DCG:RhoFRDot2}
\end{equation}
Unlike breeding by cross-glide, dislocation  multiplication by fixed
Frank-Read sources is not thermally activated process, and thus, it
remains operative even at low temperatures. At finite temperatures
however both mechanisms contribute to the total dislocation
multiplication rate. 
  
It should pointed out that both Eqs.~(\ref{DCG:RhoDot3}) and
(\ref{DCG:RhoFRDot2}) predict that $\rho^\alpha$ grows as
$(\gamma^\alpha)^2$. This rate of growth is indeed observed in
many crystals during the intermediate stages of dislocation
multiplication. 

\subsection{Attrition by pair annihilation}

Dislocation densities are often observed to attain a saturation
density at sufficiently large strains \cite{ashby:1972}. This
saturation stage arises as a result of the competition between
dislocation multiplication mechanisms such as double cross-slip
and pair annihilation. For instance, spontaneous annihilation is
observed in metals such as copper for screw dipole heights less
than 1 nm \cite{essmann:1979}. Pair annihilation is mainly the
result of the cross-slip of screw segments of opposite sign
\cite{AriasJoannopoulos1994, WangStrachanCaginGoddard2000,
RasmussenVeggeLeffers2000}.

Huang {\it et al.} \cite{HuangGhoniemDelaRubia1999}, have studied
the dynamic stability of short-range linear-elastic interactions
between two dislocations of parallel line vectors which glide on
two parallel slip planes in \bcc crystals. Here we develop a
similar but somewhat simpler linear-elastic model of dislocation
pair annihilation. Thus, our goal is to estimate the frequency
with which two parallel screw segments moving on parallel planes
will acquire converging trajectories leading to their mutual
annihilation. We note that, for simplicity, our analysis is
restricted to annihilation between pairs of dislocations belonging
to the same slip system.

Consider a screw segment moving under the action of an applied
shear stress $\tau$. The segment follows a path which brings it in
close proximity to a second immobile screw segment,
Fig.~7.  Let $(x,y)$ be coordinates centered
at the fixed dislocation such that $x$ points in the direction of
motion of the incoming dislocation and $y$ points in the direction
of the cross-slip plane, Fig.~7. The
interaction force per unit length exerted on the moving segment along 
the cross slip plane is
\cite{hirth:1968}:
\begin{equation}\label{PA:FInt}
f_y = - K \frac{y}{r^2}
\end{equation}
where $K$ is the pre-logarithmic factor for a screw segment, and
\begin{equation}\label{PA:R}
r = \sqrt{x^2 + y^2 + 2 x y \cos \alpha}
\end{equation}
where $\alpha$ is the angle subtended by the $x$ and $y$
directions, Fig.~7. For \bcc crystals,
$\alpha = 2\pi/3$. For isotropic crystals, $K = \mu b^2/4\pi$. In
view of Eq.~(\ref{Velocity}), the equation of motion for the
incoming screw dislocation is:
\begin{eqnarray}
%
\dot{y} &=& - 2 l_P \nu_D {\rm e}^{- \beta E^{\rm cross}}
\sinh\left[\beta E^{\rm cross} \frac{(K y/b r^2)}{\tau_0} \right]
\label{PA:YDot}
\end{eqnarray}
for 
$y > \kappa_0 > 0$, where the characteristic distance
\begin{equation}\label{PA:Kappa0}
\kappa_0 = \frac{K}{b\tau_0}
\end{equation}
is determined by the condition $f_y = b \tau_0$. 
For $y < \kappa_0$ dislocation motion is no longer a thermally
activated process and therefore Eq.~\ref{PA:YDot}, which is based on the
transition-state theory, does no longer apply. In this regime,
where the interaction force exceeds the Peierls barrier, dislocation
mobility increases significantly\cite{suzuki:1991}.    

Depending on the value of $y$ as $x \to -\infty$, the moving
segment bypasses or is captured by the second segment. The precise
calculation of the trajectories of the moving dislocation requires
numerical computation. However, the essence of the annihilation
mechanism can be captured by the following simple argument.
Consider a dislocation initially at rest at position $(x=0, y_0)$.
Subsequently, the motion of the dislocation for $y \ge \kappa_0$
is governed by the equation:
\begin{equation}\label{PA:YDot2}
\dot{y} = - 2 l_P \nu_D {\rm e}^{- \beta E^{\rm cross}} \sinh
\left( \beta E^{\rm cross} \frac{(K /b y)}{\tau_0} \right)
\end{equation}
This equation is separable, and the time required for the
annihilation process follows simply as:
\begin{equation}\label{PA:Tc}
t = \int_{\kappa_0}^{y_0} \left\{ 2 l_P \nu_D {\rm e}^{- \beta E^{\rm
cross}} \sinh \left( \beta E^{\rm cross} \frac{(K /b y)}{\tau_0}
\right) \right\}^{-1} dy
\end{equation}
In order to facilitate this calculation, the time required to travel
from $y = \kappa_0$ to $y = 0$ is neglected in the previous equation
due to the much higher dislocation mobility in this regime and the
small value of $\kappa_0$, of the order of few $b$ for BCC crystals.
Additionally, we may simply set
$\sinh(x) \approx x$ to a first approximation, with the result:
\begin{equation}\label{PA:Tc2}
t = \frac{{\rm e}^{\beta E^{\rm cross}} }{2 l_P \nu_D \beta E^{\rm
cross} } \frac{y_0^2 - \kappa_0^2}{2 \kappa_0}
\end{equation}
On the other hand, the time which the incoming dislocation spends
at distances of the order of $y_0$ to the receiving dislocation is
\begin{equation}\label{PA:Tc3}
t \sim \frac{b \rho^\alpha}{\dot{\gamma}^\alpha} y_0
\end{equation}
For annihilation to be possible the annihilation time
(\ref{PA:Tc2}) must be less than the time (\ref{PA:Tc3}) which the
incoming dislocation spends in the immediate vicinity of the
target dislocation. This yields the condition:
\begin{equation}\label{PA:Yc}
y_0 \leq \kappa = \kappa_0 \left( A + \sqrt{A^2 + 1} \right)
\end{equation}
where 
\begin{equation}\label{PA:A}
A = {\rm e}^{- \beta E^{\rm cross}} \beta E^{\rm cross} \dot{\gamma}_0^{\rm
cross}/\dot{\gamma}^\alpha 
\end{equation}
is a factor depending on the strain rate and temperature,
\begin{equation}\label{PA:GammaDot0}
\dot{\gamma}_0^{\rm cross} = 2 b \rho l_P \nu_D
\end{equation}
is a reference slip-strain rate, and $\kappa$ may now be regarded as
an effective pair annihilation distance.  The cut-off value
$\kappa_{c}$, corresponding to the maximal pair annihilation distance,
is the effective screening distance which can be set equal to the mean
distance between dislocations.
 
A simple expression which interpolates between the extreme values of
$\kappa$ is
\begin{equation}\label{PA:Kappa}
\frac{1}{\kappa} = \frac{1}{\kappa_c} + \frac{1}{\kappa_0 \left( A + \sqrt{A^2 + 1} \right)}
\end{equation}
It follows that the critical pair-annihilation distance $\kappa$
decreases with increasing strain rate and decreasing temperature.
Thus, at high strain rates the dislocation velocities are high and
the probability of being captured by another dislocation
diminishes accordingly. Additionally, an increase in temperature
increases the dislocation mobility and speeds up the annihilation
process, which results in an attendant increase in annihilation
rates. As will be demonstrated in
Section~\ref{ComparisonExperiment}, these trends are important in
order to capture the temperature and strain-rate dependence of the
stage I-II transition in Ta.

The rate of dislocation attrition due to pair annihilation may
finally be estimated as follows. Over a differential of time $dt$,
the number of annihilation events per unit volume is
\begin{equation}
dN^\alpha = \frac{1}{2 L^{\rm cross}}(\rho^\alpha)^2 \kappa
v^\alpha dt
\end{equation}
The attendant loss of dislocation length per unit volume is
\begin{equation}
d\rho^\alpha = - 2 L^{\rm cross} dN^\alpha
\end{equation}
Combining these relations and using Orowan's formula, the rate of
pair annihilation may finally be expressed in the form
\begin{equation}
b\dot{\rho}^\alpha = - \kappa \rho^\alpha \dot{\gamma}^\alpha
\label{DensityAnnihilation}
\end{equation}

\subsection{Dislocation multiplication rate}

Combining \eqref{DCG:RhoDot3}, \eqref{DCG:RhoFRDot2}) and
\eqref{DensityAnnihilation}, the total rate of change of the
dislocation density follows as
\begin{equation}\label{DotRho}
b \dot{\rho}^\alpha = \lambda \sqrt{\rho^\alpha}
\dot{\gamma}^\alpha - \kappa \rho^\alpha \dot{\gamma}^\alpha
\end{equation}
where $\lambda = \lambda_{\rm CS} + \lambda_{\rm FR}$. 
Evidently, $\dot{\rho}^\alpha = 0$ upon the attainment of the
saturation density:
\begin{equation}\label{RhoSat}
\rho^{\rm sat} = \left( \frac{\lambda}{\kappa} \right)^2
\end{equation}
Dividing through by $\rho^{\rm sat}$, (\ref{RhoSat}) may be recast
in the form:
\begin{equation}\label{DotRho2}
\frac{\dot{\rho}^\alpha}{\rho^{\rm sat}} = \left(
\sqrt{\frac{\rho^\alpha}{\rho^{\rm sat}}} -
\frac{\rho^\alpha}{\rho^{\rm sat}} \right)
\frac{\dot{\gamma}^\alpha}{\gamma^{\rm sat}}
\end{equation}
where
\begin{equation}
\gamma^{\rm sat} = \frac{b}{\kappa}
\end{equation}
is a saturation slip strain. Equation (\ref{DotRho2}) can be
integrated, yielding:
\begin{equation}\label{Rho}
\frac{\rho^\alpha}{\rho^{\rm sat}} = \left[
1+\left(\sqrt{\frac{\rho_0^\alpha}{\rho^{\rm sat}}}-1\right)
\exp\left(-\frac{1}{2}\frac{\gamma^\alpha}{\gamma^{\rm sat}}
\right)\right]^2
\end{equation}

The rate equation (\ref{DotRho2}) expresses a competition
between the dislocation multiplication and annihilation
mechanisms. For slip strains $\gamma^\alpha \ll \gamma^{\rm
sat}$, the multiplication term dominates and, as noted previously,
the dislocation density $\dot{\rho}^\alpha$ grows as
$(\dot{\gamma}^\alpha)^2$. By contrast, when $\gamma^\alpha
\gg \gamma^{\rm sat}$ the rates of multiplication and annihilation
balance out and saturation sets in. After saturation is attained,
the dislocation density remains essentially unchanged. It should
be carefully noted, that, in view of (\ref{PA:Kappa}), the
saturation slip strain $\gamma_{\rm sat}$ is a function of
temperature and strain rate. In particular, $\gamma_{\rm sat}$
decreases with increasing temperature and decreasing strain rate.
Since the stage I-II transition strain scales roughly with
$\gamma_{\rm sat}$, we expect these trends to be exhibited by the
transition strain itself, in accordance with experimental
observation.

\section{Comparison with experiment}
\label{ComparisonExperiment}

We proceed to validate the theory against the uniaxial tests on Ta
single crystals of Mitchell and Spitzig \cite{mitchell:1965}. In these
tests, 99.97\%-pure Ta specimens were loaded in tension along the
$[213]$ crystallographic axis, at various combinations of temperature
and strain rate. In particular we considered temperatures ranging from
$296 \; K$ to $573 \; K$, and strain rates ranging from 
$10^{-1} \; s^{-1}$ to $10^{-5} \; s^{-1}$. The
numerical procedure employed for the integration of the
constitutive equations has been described elsewhere
\cite{OrtizStainier1999}. The constitutive update is fully
implicit, with the active systems determined iteratively so as to
minimize an incremental work function. All stress-strain curves
are reported in terms of nominal stress and engineering strain.

The material property set used in calculations is collected in
Table~\ref{tab:properties}. The elastic moduli $C_{11}$, $C_{12}$
and $C_{44}$ were obtained by fitting to the tables of Simmons and
Wang \cite{simmons:1971}. The ratio between edge and screw
dislocation-line energies ($U^{\text{edge}}/U^{\text{screw}}$) is
taken from the atomistic calculations of Wang {\it et al.}
\cite{WangStrachanCaginGoddard2000}. The remaining parameters have been
obtained by fitting to the experimental data of Mitchell and
Spitzig \cite{mitchell:1965}. Note that we do not give any value for
$L^{\rm sat}$, since it has no measurable influence in the range of
temperatures considered here. It is expected, however, to play a role at
higher temperatures. 

\begin{table}
  \caption{Parameter set for Tantalum}
  \label{tab:properties}
  \smallskip
  \renewcommand{\arraystretch}{1.2}
  \centering
  \begin{tabular}{ccc}
   \hline
    Parameter           & Value                  & Units \\
   \hline
    $C_{11}$            & $266.49-0.021 T$ $({}^{\footnotesize *})$
           & [GPa] \\
    $C_{12}$            & $156.25-0.006 T$       & [GPa] \\
    $C_{44}$            & $ 90.02-0.015 T$       & [GPa] \\
   \hline
    $b$                 & $2.86\times 10^{-10}$       & [m] \\
   \hline
    $E^{\text{kink}}$   & $0.70$                 & [eV] \\
    $L^{\rm kink}/b$    & $13$                   & \\
    $U^{\text{edge}}/\mu b^{2}$ $({}^{\footnotesize **})$ & $0.216 $
            & \\
    $U^{\text{edge}}/U^{\text{screw}}$ & $1.77 $ & \\
   \hline
    $\bar{l}/b$         & $5$                    & \\
    $L^{\rm junct}/b$   & $20$                   & \\
   \hline
    $E^{\text{cross}}$  & $0.67$                 & [eV] \\
    $L^{\rm cross}/b$   & $13$                   & \\
    $\kappa_c/b$        & $1250$                 & \\
    $\lambda_{\rm FR}$  & $2.3$                  & \\
   \hline
    $\rho_{0}$          & $10^{12}$              & [m$^{-2}$] \\
    $g_0$               & $8.0$                  & [MPa]\\
   \hline
    $a_{0}$             & $0.01$                 &  \\
   \hline
    \multicolumn{3}{l}{\rule{0pt}{2.5ex} ${}^{\footnotesize *}$
        $T$ in Kelvin.}
         \\
    \multicolumn{3}{l}{\rule{0pt}{2.5ex} ${}^{\footnotesize **}$
         $\mu=\frac{3}{5} C_{44} + \frac{1}{5} (C_{11}-C_{12})$.}
         \\
  \end{tabular}
\end{table}

Figs.~8 and 9 show
the predicted and measured stress-strain curves for a $[213]$ Ta
crystal over a range of temperatures and strain rates. It is
evident from these figures that the model captures salient
features of the behavior of Ta crystals such as: the dependence of
the initial yield point on temperature and strain rate; the
presence of a marked stage I of easy glide, specially at low
temperature and high strain rates; the sharp onset of stage II
hardening and its tendency to shift towards lower strains, and
eventually disappear, as the temperature increases or the strain
rate decreases; the parabolic stage II hardening at low strain
rates or high temperatures; the stage II softening at high strain
rates or low temperatures; the trend towards saturation at high
strains; and the temperature and strain-rate dependence of the
saturation stress. 

The theory reveals useful insights into the mechanisms underlying
these behaviors. For instance, since during state I the crystal
deforms in single slip and the secondary dislocation densities are
low, the Peierls resistance dominates and the temperature and
strain-rate dependency of yield owes mainly to the thermally
activated formation of kinks and crossing of forest dislocations.
It is interesting to note that during this stage the effect of
increasing (decreasing) temperature is similar to the effect of
decreasing (increasing) strain rate, as noted by Tang {\it et al.}
\cite{TangDevincreKubin1999}. The onset of stage II is due to the
activation of secondary systems. The rate at which these secondary
systems harden during stage I depends on the rate of dislocation
multiplication in the primary system. This rate is in turn
sensitive to the saturation strain $\gamma^{\rm sat}$, which
increases with strain rate and decreases with temperature. As a
result, the length of the stage I of hardening is predicted to
increase with strain rate and decrease with temperature, as
observed experimentally. Finally, the saturation stress is mainly
governed by the forest hardening mechanism and, in particular, by
the strength of the forest obstacles. This process is less
thermally activated than the Peierls stress, since the
corresponding energy barriers are comparatively higher.
Consequently, the stress-strain curves tend to converge in this
regime, in keeping with observation.

The apparent softening observed in simulation results at the lowest
temperature ($296\;K$) and the highest strain rate ($10^{-1}\;s^{-1}$)
is actually an effect of the boundary conditions, which allow for some level
of rotation of the specimen. Since in those cases, the material
hardening is relatively low (stage I only), this geometrical softening
dominates the apparent macroscopic behavior. In the other cases,
the activation of several systems limits the extend of
the rotations reducing the effects of the macroscopic hardening.
In order to simulate more precisely the experimental boundary
conditions, a model of the entire specimen allowing
for a non-homogeneous deformation field should be considered.

The effect of temperature and strain-rate on hardening
is also illustrated by the evolution of slip activity in the
primary and secondary slip systems, $D1$ and $D4''$, respectively.
Figs.~10 and 11 show the evolution of
slip strains and dislocation densities as a function of
temperature. It is evident from these figures that dislocation
density saturation in the primary system occurs earlier as the
temperature is increased, resulting in the activation of the
secondary system and in the onset of stage II at lower strains. A
similar effect is observed when the strain rate is decreased,
Figs.~12 and 13.

The effect of loading direction on the hardening rate is
illustrated in Fig.~14. The figure shows the
stress-strain curves obtained by loading the crystal in the [213],
[101] and [111] directions at a strain rate of $10^{-3}$
s${}^{-1}$ and a temperature of 373 K. Each of these loading
directions results in the activation of a different set of slip
systems. As may be observed in the figure, the higher initial
yield stress in the [111] direction relative to the baseline [213]
direction, and the initial negative hardening rate, are fairly well
captured by the model. The experimental curve exhibits a
subsequent upturn, most likely due to the activation of additional
secondary systems, which is not captured by the model. Loading in
the [101] direction results in the activation of a large number of
systems from the outset and the rate of hardening is
correspondingly high. The model appears to over-predict the rate
of hardening.

However, as was noted in \cite{cuitino:1992}, the stress-strain
curve for crystals loaded in high-symmetry orientations are
extremely sensitive to small misalignments in the loading axis,
which accounts for the large experimental scatter characteristic
of those orientations. This extreme sensitivity is due to the fact
that small deviations from a high-symmetry loading axis, of the
order of a degree or less, may result in the activation of a
different set of slip systems, which may in turn have a large
effect on the hardening rates. This pathological behavior is
illustrated in Fig.~15, which shows the
stress-strain curve obtained by randomly offsetting less than one degree 
the nominal loading axis [101]. The resulting
reduction in the hardening rate is quite remarkable. In addition,
the experimental curve falls between the bounds of the
stress-strain curves for the nominal and perturbed directions.
This comparison exemplifies the need to take experimental scatter
into account when assessing the fidelity of models, specially
where crystals loaded along directions of high symmetry are
concerned.

\section{Summary and conclusions}

We have developed a micromechanical model of the hardening,
rate-sensitivity and thermal softening of \bcc crystals. The model
is predicated upon the consideration of an `irreducible' set of
unit processes, consisting of: double-kink formation and thermally
activated motion of kinks; the close-range interactions between
primary and forest dislocation, leading to the formation of jogs;
the percolation motion of dislocations through a random array of
forest dislocations introducing short-range obstacles of different
strengths; dislocation multiplication due to breeding by double
cross-slip; and dislocation pair-annihilation. Each of these
processes accounts for--and is needed for matching--salient and
clearly recognizable features of the experimental record. In
particular, on the basis of detailed comparisons with the
experimental data of Mitchell and Spitzig \cite{mitchell:1965},
the model is found to capture: the dependence of the initial yield
point on temperature and strain rate; the presence of a marked
stage I of easy glide, specially at low temperature and high
strain rates; the sharp onset of stage II hardening and its
tendency to shift towards lower strains as the temperature
increases or the strain rate decreases; the initial parabolic
hardening followed by saturation within the stage II of hardening;
the temperature and strain-rate dependence of the saturation
stress; and the orientation dependence of the hardening rates.

The choice of analysis tools which we have brought to bear on the
unit processes of interest, e.~g., transition-state theory,
stochastic modeling, and simple linear-elastic models of defects
and their interactions, is to a large extent conditioned by our
desire to derive closed-form analytical expressions for all
constitutive relations. As noted throughout the paper, many of the
mechanisms under consideration are amenable to a more complete
analysis by recourse to atomistic or continuum methods. However,
at this stage of development, direct simulation methods, be it
atomistic or continuum based, tend to produce unmanageable
quantities of numerical data and rarely result in analytical
descriptions of effective behavior. The daunting task of
post-processing these data sets and uncovering patterns and laws
within them which can be given analytical expression is as yet a
largely unfulfilled goal of multiscale modeling.

This larger picture notwithstanding, one concrete and workable
link between micromechanical models and first-principles
calculations concerns the calculation of material constants. A
partial list relevant to the present model includes: energy
barriers and attempt frequencies for double-kink formation, kink
migration, dislocation unpinning, cross-slip, and pair
annihilation; dislocation-line and jog energies; and junction
strengths.  Other properties which have yielded to direct
calculation include the volumetric equation of state (EoS), the
pressure dependence of yield, and the pressure and temperature
dependence of elastic moduli. References to recent work concerned
with the calculation of these material properties have been given
throughout the paper. As noted earlier, these results provide a
suitable basis for future extensions of the present model to
higher temperatures, pressures and strain-rates.

\section*{Acknowledgments}
The support of the DOE through Caltech's ASCI Center for the
Simulation of the Dynamic Response of Materials is gratefully
acknowledged. LS also wishes to acknowledge support from the
Belgian National Fund for Scientific Research (FNRS).

\bibliographystyle{unsrt}

\begin{thebibliography}{10}

\bibitem{BulatovKubin1998}
{V.V.} Bulatov and {L.P.} Kubin.
\newblock Dislocation modelling at atomistic and mesoscopic scales.
\newblock {\em Current Opinion in Solid State \& Materials Science},
  3(6):558--561, 1998.

\bibitem{Phillips1998}
R.~Phillips.
\newblock Multiscale modeling in the mechanics of materials.
\newblock {\em Current Opinion in Solid State \& Materials Science},
  3(6):526--532, 1998.

\bibitem{CampbellFoilesHuang1998}
{G.H.} Campbell, {S.M.} Foiles, {H.C.} Huang, {D.A.} Hughes, {W.E.} King,
  {D.H.} Lassila, {D.J.} Nikkel, {T.D.} de~la Rubia, {J.Y.} Shu, and {V.P.}
  Smyshlyaev.
\newblock Multi-scale modeling of polycrystal plasticity: a workshop report.
\newblock {\em Materials Science and Engineering A-Structural Materials
  Properties Microstructure and Processing}, 251(1-2):1--22, 1998.

\bibitem{PhillipsRodneyShenoy1999}
R.~Phillips, D.~Rodney, V.~Shenoy, E.~Tadmor, and M.~Ortiz.
\newblock Hierarchical models of plasticity: dislocation nucleation and
  interaction.
\newblock {\em Modelling and Simulation in Materials Science and Engineering},
  7(5):769--780, 1999.

\bibitem{MoriartyXuSoderlind1999}
{J.A.} Moriarty, W.~Xu, P.~Soderlind, J.~Belak, {L.H.} Yang, and J.~Zhu.
\newblock Atomistic simulations for multiscale modeling in bcc metals.
\newblock {\em Journal of Engineering Materials and Technology- Transactions of
  the ASME}, 121(2):120--125, 1999.

\bibitem{Baskes1999}
{M.I.} Baskes.
\newblock The status role of modeling and simulation in materials science and
  engineering.
\newblock {\em Current Opinion in Solid State \& Materials Science},
  4(3):273--277, 1999.

\bibitem{cuitino:1992}
A.~M. Cuiti\~no and M.~Ortiz.
\newblock Computational modelling of single crystals.
\newblock {\em Modelling and Simulation in Materials Science and Engineering},
  1:255--263, 1992.

\bibitem{HuangGhoniemDelaRubia1999}
{H.C.} Huang, N.~Ghoniem, {T.D.} de~la Rubia, M.~Rhee, H.~Zbib, and J.~Hirth.
\newblock Stability of dislocation short-range reactions in bcc crystals.
\newblock {\em Journal of Engineering Materials and Technology- Transactions of
  the ASME}, 121(2):143--150, 1999.

\bibitem{mitchell:1965}
T.E. Mitchell and W.A. Spitzig.
\newblock Three-stage hardening in tantalum single crystals.
\newblock {\em Acta Metallurgica}, 13:1169--1179, 1965.

\bibitem{Lee1969}
E.~H. Lee.
\newblock Elastic-plastic deformation at finite strains.
\newblock {\em Journal of Applied Mechanics}, 36:1, 1969.

\bibitem{Teodosiu1970}
C.~Teodosiu.
\newblock A dynamic theory of dislocations and its applications to the theory
  of the elastic-plastic continuum.
\newblock In J.~A. Simmons, editor, {\em Conf. Fundamental Aspects of
  Dislocation Theory}, volume~2, page 837, Washington, 1969. Natl. Bureau of
  Standards Special Publication.

\bibitem{AsaroRice1977}
R.~J. Asaro and J.~R. Rice.
\newblock Strain localization in ductile single crystals.
\newblock {\em Journal of the Mechanics and Physics of Solids}, 25:309, 1977.

\bibitem{Havner1973}
K.~S. Havner.
\newblock On the mechanics of crystalline solids.
\newblock {\em Journal of the Mechanics and Physics of Solids}, 21:383, 1973.

\bibitem{HillRice1972}
R.~Hill and J.~R. Rice.
\newblock Constitutive analysis of elastic-plastic crystals at arbitary
  strains.
\newblock {\em Journal of the Mechanics and Physics of Solids}, 20:401, 1972.

\bibitem{Rice1971}
J.~R. Rice.
\newblock Inelastic constitutive relations for solids: an internal-variable
  theory and its applications to metal plasticity.
\newblock {\em Journal of the Mechanics and Physics of Solids}, 19:433, 1971.

\bibitem{WassermanStixrudeCohen1996}
E.~Wasserman, L.~Stixrude, and {R.E.} Cohen.
\newblock Thermal properties of iron at high pressures and temperatures.
\newblock {\em Physical Review B-Condensed Matter}, 53(13):8296--8309, 1996.

\bibitem{CohenStixrudeWasserman1997}
{R.E.} Cohen, L.~Stixrude, and E.~Wasserman.
\newblock Tight-binding computations of elastic anisotropy of fe, xe, and si
  under compression.
\newblock {\em Physical Review B-Condensed Matter}, 56(14):8575--8589, 1997.

\bibitem{SoderlindMoriarty1998}
P.~Soderlind and {J.A.} Moriarty.
\newblock First-principles theory of ta up to 10 mbar pressure: Structural and
  mechanical properties.
\newblock {\em Physical Review B-Condensed Matter}, 57(17):10340--10350, 1998.

\bibitem{Steinle-neumannStixrudeCohen1999}
G.~Steinle-Neumann, L.~Stixrude, and {R.E.} Cohen.
\newblock First-principles elastic constants for the hcp transition metals fe,
  co, and re at high pressure.
\newblock {\em Physical Review B-Condensed Matter}, 60(2):791--799, 1999.

\bibitem{BulatovRichmondGlazov1999}
{V.V.} Bulatov, O.~Richmond, and {M.V.} Glazov.
\newblock An atomistic dislocation mechanism of pressure- dependent plastic
  flow in aluminum.
\newblock {\em Acta Materialia}, 47(12):3507--3514, 1999.

\bibitem{CohenGulserenHemley2000}
{R.E.} Cohen, O.~Gulseren, and {R.J.} Hemley.
\newblock Accuracy of equation-of-state formulations.
\newblock {\em American Mineralogist}, 85(2):338--344, 2000.

\bibitem{simmons:1971}
G.~Simmons and H.~Wang.
\newblock {\em Single crystal elastic constants and calculated aggregate
  properties: A handbook}.
\newblock M.I.T. Press, Cambridge, Massachusetts, 1971.

\bibitem{weiner:1983}
J.~H. Weiner.
\newblock {\em Statistical Mechanics of Elasticity}.
\newblock John Wiley \& Sons, New York, 1983.

\bibitem{hull:1984}
D.~Hull and D.~J. Bacon.
\newblock {\em Introduction to Dislocations}, volume~37 of {\em International
  Series on Materials Science and Technology}.
\newblock Elsevier Science Inc., 3rd edition, 1984.

\bibitem{BengusDolginTabachnikova1985}
{V.Z.} Bengus, {A.M.} Dolgin, {E.D.} Tabachnikova, and {Y.V.} Efimov.
\newblock Twinning asymmetry of niobium single-crystals at 4.2 k.
\newblock {\em Physics of Metals}, 5(5):992--998, 1985.

\bibitem{SeegerHollang2000}
A.~Seeger and L.~Hollang.
\newblock The flow-stress asymmetry of ultra-pure molybdenum single crystals.
\newblock {\em Materials Transactions JIM}, 41(1):141--151, 2000.

\bibitem{franciosi:1983}
P.~Franciosi and A.~Zaoui.
\newblock Glide mechanisms in b.c.c. crystals: an investigation of the case of
  $\alpha$-iron through multislip and latent hardening tests.
\newblock {\em Acta Metallurgica}, 31:1331, 1983.

\bibitem{OrtizStainier1999}
M.~Ortiz and L.~Stainier.
\newblock The variational formulation of viscoplastic constitutive updates.
\newblock {\em Computer Methods in Applied Mechanics and Engineering},
  171(3-4):419--444, 1999.

\bibitem{DuesberyVitekBowen1973}
Duesbery, V.~Vitek, and Bowen.
\newblock {\em Proceedings of the Royal Society of London}, A332:85, 1973.

\bibitem{Vitek1976}
V.~Vitek.
\newblock {\em Proceedings of the Royal Society of London}, A352:109, 1976.

\bibitem{Vitek1992}
V.~Vitek.
\newblock Structure of dislocation cores in metallic materials and its impact
  on their plastic behavior.
\newblock {\em Progress in Material Science}, 36:1--27, 1992.

\bibitem{XuMoriarty1996}
W.~Xu and {J.A.} Moriarty.
\newblock Atomistic simulation of ideal shear strength, point defects, and
  screw dislocations in bcc transition metals: Mo as a prototype.
\newblock {\em Physical Review B-Condensed Matter}, 54(10):6941--6951, 1996.

\bibitem{DuesberyVitek1998}
{M.S.} Duesbery and V.~Vitek.
\newblock Plastic anisotropy in bcc transition metals.
\newblock {\em Acta Materialia}, 46(5):1481--1492, 1998.

\bibitem{Ismail-beigiArias2000}
S.~Ismail-Beigi and {T.A.} Arias.
\newblock Ab initio study of screw dislocations in mo and ta: A new picture of
  plasticity in bcc transition metals.
\newblock {\em Physical Review Letters}, 84(7):1499--1502, 2000.

\bibitem{WangStrachanCaginGoddard2000}
G.~Wang, A.~Strachan, T.~Cagin, and W.A.~III Goddard.
\newblock Molecular dynamics simulations of 1/2a$<111>$ screw dislocation in
  ta.
\newblock {\em Materials Science and Engineering A}, 2000.

\bibitem{DuesberyXu1998}
{M.S.} Duesbery and W.~Xu.
\newblock The motion of edge dislocations in body-centered cubic metals.
\newblock {\em Scripta Materialia}, 39(3):283--287, 1998.

\bibitem{Hirsch1960}
P.~B. Hirsch.
\newblock In {\em 5th International Conference on Crystallography}, page 139.
  Cambridge University, 1960.

\bibitem{SeegerSchiller1962}
A.~Seeger and P.~Schiller.
\newblock {\em Acta Metallurgica}, 10:348, 1962.

\bibitem{hirth:1968}
J.~P. Hirth and J.~Lothe.
\newblock {\em {Theory of Dislocations}}.
\newblock McGraw-Hill, New York, 1968.

\bibitem{HirthHoagland1993}
{J.P.} Hirth and {R.G.} Hoagland.
\newblock Nonlinearities in the static energetics and in the kinematics of
  dislocations.
\newblock {\em Physica D}, 66(1-2):71--77, 1993.

\bibitem{XuMoriarty1998}
W.~Xu and {J.A.} Moriarty.
\newblock Accurate atomistic simulations of the peierls barrier and kink-pair
  formation energy for $<111>$ screw dislocations in bcc mo.
\newblock {\em Computational Materials Science}, 9(3-4):348--356, 1998.

\bibitem{KocksArgonAshby1975}
U.~F. Kocks, A.~S. Argon, and M.~F. Ashby.
\newblock In B.~Chalmers, J.~W. Christian, and T.~B. Massalski, editors, {\em
  Thermodynamics and Kinetics of Slip}, volume~19 of {\em Progress in Materials
  Science}. Pergamon Press, 1975.

\bibitem{TangKubinCanova1998}
M.~Tang, {L.P.} Kubin, and {G.R.} Canova.
\newblock Dislocation mobility and the mechanical response of bcc single
  crystals: A mesoscopic approach.
\newblock {\em Acta Materialia}, 46(9):3221--3235, 1998.

\bibitem{Wasserbach1986}
W.~Wasserb\"ach.
\newblock {\em Philosophical Magazine}, A53:335, 1986.

\bibitem{LachenmannShultz1970}
R.~Lachenmann and H.~Schultz.
\newblock {\em Scripta Metallurgica}, 4:33, 1970.

\bibitem{TangDevincreKubin1999}
M.~Tang, B.~Devincre, and {L.P.} Kubin.
\newblock Simulation and modelling of forest hardening in body centre cubic
  crystals at low temperature.
\newblock {\em Modelling and Simulation in Materials Science and Engineering},
  7(5):893--908, 1999.

\bibitem{suzuki:1991}
T.~Suzuki, S.~Takeuchi, and H.~Yoshinaga.
\newblock {\em Dislocation Dynamics and Plasticity}.
\newblock Springer-Verlag, 1991.

\bibitem{brailsford:1969}
A.D. Brailsford.
\newblock Electronic component of dislocation drag in metals.
\newblock {\em Physical Review}, 186:959--961, 1969.

\bibitem{baskes:1998}
{M.I.} Baskes, {R.G.} Hoagland, and T.~Tsuji.
\newblock An atomistic study of the strength of an extended-dislocation
  barrier.
\newblock {\em Modelling and Simulation in Materials Science and Engineering},
  6(1):9--18, 1998.

\bibitem{RodneyPhillips1999}
D.~Rodney and R.~Phillips.
\newblock Structure and strength of dislocation junctions: An atomic level
  analysis.
\newblock {\em Physical Review Letters}, 82(8):1704--1707, 1999.

\bibitem{ShenoyKuktaPhillips2000}
{V.B.} Shenoy, {R.V.} Kukta, and R.~Phillips.
\newblock Mesoscopic analysis of structure and strength of dislocation
  junctions in fcc metals.
\newblock {\em Physical Review Letters}, 84(7):1491--1494, 2000.

\bibitem{DannaBenoit1993}
G.~Danna and W.~Benoit.
\newblock Dynamic recovery of the microstructure of screw dislocations in
  high-purity bcc metals.
\newblock {\em Materials Science and Engineering A-Structural Materials
  Properties Microstructure and Processing}, 164(1-2):191--195, 1993.

\bibitem{RheeZbibHirth1998}
M.~Rhee, {H.M.} Zbib, {J.P.} Hirth, H.~Huang, and T.~de~la Rubia.
\newblock Models for long-/short-range interactions and cross slip in 3d
  dislocation simulation of bcc single crystals.
\newblock {\em Modelling and Simulation in Materials Science and Engineering},
  6(4):467--492, 1998.

\bibitem{KubinDevincreTang1998}
L.~P. Kubin, B.~Devincre, and M.~Tang.
\newblock {\em Journal of Computer Aided Material Design}, 5:31, 1998.

\bibitem{ZbibDeLaRubiaRhee2000}
{H.M.} Zbib, {T.D.} de~la Rubia, M.~Rhee, and {J.P.} Hirth.
\newblock 3d dislocation dynamics: stress-strain behavior and hardening
  mechanisms in fcc and bcc metals.
\newblock {\em Journal of Nuclear Materials}, 276:154--165, 2000.

\bibitem{foreman:1966}
A.~J.~E. Foreman and M.~J. Makin.
\newblock Dislocation movement through random arrays of obstacles.
\newblock {\em Philosophical Magazine}, 14:911, 1966.

\bibitem{foreman:1967}
A.~J.~E. Foreman and M.~J. Makin.
\newblock Dislocation movement through random arrays of obstacles.
\newblock {\em Canadian Journal of Physics}, 45:273, 1967.

\bibitem{kocks:1966}
U.~F. Kocks.
\newblock A statistical theory of flow stress and work-hardening.
\newblock {\em Philosophical Magazine}, 13:541, 1966.

\bibitem{ortiz:1982}
M.~Ortiz and E.~P. Popov.
\newblock A statistical theory of polycrystalline plasticity.
\newblock {\em Proceedings of the Royal Society of London}, A379:439--458,
  1982.

\bibitem{Ortiz1999}
M.~Ortiz.
\newblock Plastic yielding as a phase transition.
\newblock {\em Journal of Applied Mechanics-Transactions of the ASME},
  66(2):289--298, 1999.

\bibitem{Cuitino2000}
A.~M. Cuiti\~no, M.~Koslowski, M.~Ortiz, and L.~Stainier.
\newblock A phase-field theory of dislocation dynamics, strain hardening and
  hysteresis in ductile single crystals at low temperatures.
\newblock Philosophical Magazine, submitted for publication, 2000.

\bibitem{ChangBulatovYip1999}
{J.P.} Chang, {V.V.} Bulatov, and S.~Yip.
\newblock Molecular dynamics study of edge dislocation motion in a bcc metal.
\newblock {\em Journal of Computer-Aided Materials Design}, 6(2-3):165--173,
  1999.

\bibitem{mughrabi:1975}
H.~Mughrabi.
\newblock {Description of the Dislocation Structure after Unidirectional
  Deformation at Low Temperatures}.
\newblock In A.S. Argon, editor, {\em {Constitutive Equations in Plasticity}},
  pages 199--250, Cambridge, Mass, 1975. MIT Press.

\bibitem{grosskreutz:1975}
J.~C. Grosskreutz and H.~Mughrabi.
\newblock {Description of the work-hardened structure at low temperature in
  cyclic deformation.}
\newblock In A.~S. Argon, editor, {\em {Constitutive Equations in Plasticity}},
  pages 251--326, Cambridge, Mass, 1975. MIT Press.

\bibitem{cuitino:1993}
A.~M. Cuiti\~no and M.~Ortiz.
\newblock Constitutive modeling of ${\rm l1_2}$ intermetallic crystals.
\newblock {\em Materials Science and Engineering}, A170:111--123, 1993.

\bibitem{franciosi:1982}
P.~Franciosi and A.~Zaoui.
\newblock Multislip in f.c.c. crystals: A theoretical approach compared with
  experimental data.
\newblock {\em Acta Metallurgica}, 30:1627, 1982.

\bibitem{LeeSubhashRavichandran1999}
{Y.J.} Lee, G.~Subhash, and G.~Ravichandran.
\newblock Constitutive modeling of textured body-centered- cubic (bcc)
  polycrystals.
\newblock {\em International Journal of Plasticity}, 15(6):625--645, 1999.

\bibitem{kuhlmann:1989}
D.~Kuhlmann-Wilsdorf.
\newblock Theory of plastic deformation: properties of low energy dislocation
  structures.
\newblock {\em Materials Science and Engineering}, A113:1, 1989.

\bibitem{johnston:1959}
W.~G. Johnston and J.~J. Gilman.
\newblock Dislocation velocities, dislocation densities and plastic flow in
  lithium fluoride crystals.
\newblock {\em Journal of Applied Physics}, 30:129, 1959.

\bibitem{johnston:1960}
W.~G. Johnston and J.~J. Gilman.
\newblock Dislocation multiplication in lithium fluoride crystals.
\newblock {\em Journal of Applied Physics}, 31:632, 1960.

\bibitem{gillis:1965}
P.~P. Gillis and J.~Gilman.
\newblock {Dynamical Dislocation Theory of Crystal Plasticity. II. Easy Glide
  and Strain Hardening}.
\newblock {\em Journal of Applied Physics}, 36:3380, 1965.

\bibitem{essmann:1973}
U.~Essmann and M.~Rapp.
\newblock Slip in copper crystals following weak neutron bombardment.
\newblock {\em Acta Metallurgica}, 21:1305, 1973.

\bibitem{Lagerlof1993}
{K.P.D.} Lagerlof.
\newblock On deformation twinning in bcc metals.
\newblock {\em Acta Metallurgica et Materialia}, 41(7):2143--2151, 1993.

\bibitem{Dybiec1995}
H.~Dybiec.
\newblock Model of the early deformation stage of bcc metals in
  low-temperature.
\newblock {\em Zeitschrift f\"ur Metallkunde}, 86(7):512--517, 1995.

\bibitem{RasmussenJacobsenLeffers1997}
T.~Rasmussen, {K.W.} Jacobsen, T.~Leffers, {O.B.} Pedersen, {S.G.} Srinivasan,
  and H.~Jonsson.
\newblock Atomistic determination of cross-slip pathway and energetics.
\newblock {\em Physical Review Letters}, 79(19):3676--3679, 1997.

\bibitem{RaoParthasarathyWoodward1999}
S.~Rao, {T.A.} Parthasarathy, and C.~Woodward.
\newblock Atomistic simulation of cross-slip processes in model fcc structures.
\newblock {\em Philosophical Magazine A-Physics of Condensed Matter Structure
  Defects and Mechanical Properties}, 79(5):1167--1192, 1999.

\bibitem{li:1961}
J.~C.~M. Li.
\newblock Cross slip and cross climb induced by a locked dislocation.
\newblock {\em Journal of Applied Physics}, 32:593, 1961.

\bibitem{FrankRead1950}
F.C. Frank and W.T. Read.
\newblock In {\em Symposium on Plastic Deformation of Crystalline Solids},
  page~44, Pittsburgh, 1950. Carnegie Institute of Technology.

\bibitem{ashby:1972}
M.~F. Ashby.
\newblock The deformation of plastically non-homogeneous alloys.
\newblock In A.~Kelly and R.~B. Nicholson, editors, {\em Strengthening Methods
  in Crystals}. Wiley, 1972.

\bibitem{essmann:1979}
U.~Essmann and H.~Mughrabi.
\newblock Annihilation of dislocations during tensile and cyclic deformation
  and limits of dislocation densities.
\newblock {\em Philosophical Magazine}, A 40:731--756, 1979.

\bibitem{AriasJoannopoulos1994}
{T.A.} Arias and {J.D.} Joannopoulos.
\newblock Ab-initio theory of dislocation interactions - from close-range
  spontaneous annihilation to the long- range continuum-limit.
\newblock {\em Physical Review Letters}, 73(5):680--683, 1994.

\bibitem{RasmussenVeggeLeffers2000}
T.~Rasmussen, T.~Vegge, T.~Leffers, {O.B.} Pedersen, and {K.W.} Jacobsen.
\newblock Simulation of structure and annihilation of screw dislocation
  dipoles.
\newblock {\em Philosophical Magazine A-Physics of Condensed Matter Structure
  Defects and Mechanical Properties}, 80(5):1273--1290, 2000.

\end{thebibliography}

\section*{List of Figures}
\begin{description}
\item[Figure 1:] Schematic of the double-kink mechanism.
\item[Figure 2:] Temperature dependence of the effective Peierls stress
for various strain rates. Note that the typical order of magnitude
of $\dot{\gamma}^{\rm kink}_0 = 10^6\:s^{-1}$.
\item[Figure 3:]Bow-out mechanism for a dislocation segment bypassing an
obstacle pair.
\item[Figure 4:]Schematic of energy variation as a function of a
reaction coordinate during dislocation intersection and crossing. 
\item[Figure 5:]Schematic of jog formation during dislocation
intersection. 
\item[Figure 6:]Schematic of a dislocation line overcoming a junction.
\item[Figure 7:]Interaction between two parallel screw dislocations of
opposite sign moving on parallel slip planes. 
\item[Figure 8:] Temperature dependence of stress-strain curves for
  $[213]$ Ta single crystal ($\dot{\epsilon}=10^{-3}$ s${}^{-1}$).(a)
  Experimental data of Mitchell \& Spitzig \cite{mitchell:1965}. (b)
  Predictions of the model.
\item[Figure 9:]Strain-rate dependence of stress-strain curves for
  $[213]$ Ta single crystal ($T$=373 K). (a)
  Experimental data of Mitchell \& Spitzig \cite{mitchell:1965}. (b)
  Predictions of the model.
\item[Figure 10:] Slip activity as a function of temperature
  ($\dot{\epsilon}=10^{-3}$ s${}^{-1}$).
\item[Figure 11:]Evolution of dislocation densities as a function
  of temperature ($\dot{\epsilon}=10^{-3}$ s${}^{-1}$). 
\item[Figure 12:] Slip activity as a function of strain-rate ($T$=373 K).
\item[Figure 13:] Evolution of dislocation densities as a function of
  strain-rate ($T$=373 K).
\item[Figure 14:] Orientation dependence of stress-strain curves: comparison
           with experimental results
           (Mitchell \& Spitzig \cite{mitchell:1965}).
\item[Figure 15:] Effect of a small perturbation in the tensile orientation
           on the stress-strain curve.
\end{description}

\end{document}